\documentclass[amsmath,amssymb,aps,twocolumn,floats,superscriptaddress]{revtex4-2}

\usepackage{graphicx}
\usepackage{amssymb}
\usepackage{amsmath}
\usepackage{caption}
\usepackage{placeins}
\usepackage{hyperref}
\usepackage{footnote}
\usepackage{color}
\usepackage{multirow}
\usepackage{enumerate}
\usepackage{soul}
\usepackage{xcolor}
\hypersetup{
  colorlinks   = true, 
  urlcolor     = magenta, 
  linkcolor    = magenta, 
  citecolor   = magenta 
}

\newcommand{\new}[1]{{\color{black} #1}}
\newcommand{\corr}[1]{{\color{black} #1}}

\pdfoutput=1 
\begin{document}
 
\title{Quantum-impurity sensing of altermagnetic order} 

\author{V. A.S.V. Bittencourt}
\email{sant@unistra.fr}
\affiliation{Institut de Science et d’Ingénierie Supramoléculaires (ISIS, UMR7006), Universit\'{e} de Strasbourg, 67000 Strasbourg, France}  
\author{Hossein Hosseinabadi}
\affiliation{Institut f\"{u}r Physik, Johannes Gutenberg Universit\"{a}t Mainz, D-55099 Mainz, Germany}  
\author{Jairo Sinova} 
\affiliation{Institut f\"{u}r Physik, Johannes Gutenberg Universit\"{a}t Mainz, D-55099 Mainz, Germany}
\affiliation{Department of Physics, Texas AM University, College Station, Texas 77843-4242, USA}
\author{Libor \v{S}mejkal} 
\affiliation{Max Planck Institute for the Physics of Complex Systems, N\"{o}thnitzer Str. 38, 01187 Dresden, Germany}
\affiliation{Max Planck Institute for Chemical Physics of Solids, N\"{o}thnitzer Str. 40, 01187 Dresden, Germany}
\affiliation{Institute of Physics, Czech Academy of Sciences, Cukrovarnicka 10, 16200, Praha 6, Czech Republic}
\affiliation{Institut f\"{u}r Physik, Johannes Gutenberg Universit\"{a}t Mainz, D-55099 Mainz, Germany}
\author{Jamir Marino}
\affiliation{Institut f\"{u}r Physik, Johannes Gutenberg Universit\"{a}t Mainz, D-55099 Mainz, Germany}
\affiliation{Department of Physics, The State University of New York at Buffalo, NY 14260, USA}
\date{\today}
\begin{abstract}
Quantum sensing with individual spin defects has emerged as a versatile platform to probe microscopic properties of condensed matter systems. Here we demonstrate that quantum relaxometry with nitrogen-vacancy (NV) centers in diamond can reveal the anisotropic spin dynamics of altermagnetic insulators together with their characteristic spin polarised bands. We show that the distance and orientation dependent relaxation rate of a nearby quantum impurity encodes signatures of momentum space anisotropy in the spin diffusion response, a hallmark of altermagnetic order. This directional sensitivity is unprecedented in the landscape of quantum materials sensing, and it enables the distinction of altermagnets from conventional antiferromagnets via local, noninvasive measurements.
Our results could spark new NV-sensing experiments on spin transport and symmetry breaking in altermagnets, and highlight the role of NV orientation to probe anisotropic phenomena in condensed matter systems.
\end{abstract}

\maketitle

\textit{Introduction  --} Altermagnetism is a recently identified magnetic phase characterized by a unique spin arrangement of electrons in solids~\cite{PhysRevX.12.031042,PhysRevX.12.040501,Jungwirth_2025_altermagnetism,mazin_2022_aletmagnetism}. Like antiferromagnets (AFMs), altermagnets (ALMs) exhibit vanishing net magnetization. However, unlike AFMs, ALMs exhibit time-reversal broken and spin-polarised bands, a characteristic of ferromagnets. Unlike the s-wave-like isotropic spin polarisation in ferromagnets, the spin polarisation in ALMs takes d, g, or i-wave form, i.e. a momentum-space structure of spin-split bands with energy iso-surfaces that are anisotropic, and intersect at 2, 4, or 6 spin-degenerate nodes~\cite{PhysRevX.12.031042}.
Altermagnetic order was proposed in a wide range of materials~\cite{PhysRevX.12.040501,bai_2024_altermagnetism}, including metals and insulators, and two dimensional systems~\cite{mazin2023induced}, and it has been experimental confirmed in a few materials~\cite{krempasky2024altermagnetic,fedchenko2024observation, reimers2024direct,lee_2024_broken} and in some cases remains stable under ambient conditions~\cite{reimers2024direct,Jiang2025,Zhang2025}. Its discovery motivated by predicition and observation of unconventional crystal anomalous Hall effect~\cite{libor_2020,mazin_2021_prediction,Feng2022,reichlova2024observation} opens up exciting research avenues~\cite{PhysRevX.12.031042,PhysRevX.12.040501,bai_2024_altermagnetism,Jungwirth_2025_altermagnetism}, including  magnetoresistance effects \cite{smejkal_2022_giant_tunneling}, topological \cite{mazin2023induced,PhysRevLett.134.096703,Gonzalez2025Dual}, multiferroic and ferroelectric properties \cite{PhysRevLett.134.106801,libormultiferroics2024,PhysRevLett.134.106802, Duan2025Antiferroelectric,Zhu2025Emergent,Zhu2025Twodim}, spin and exciton dynamics \cite{denisov2024anisotropic, Cao2025Symmetry}, and interplay with superconductivity \cite{mazin2022notes}, to name a few.

\begin{figure}

\centering
  \includegraphics[width=1. \columnwidth]{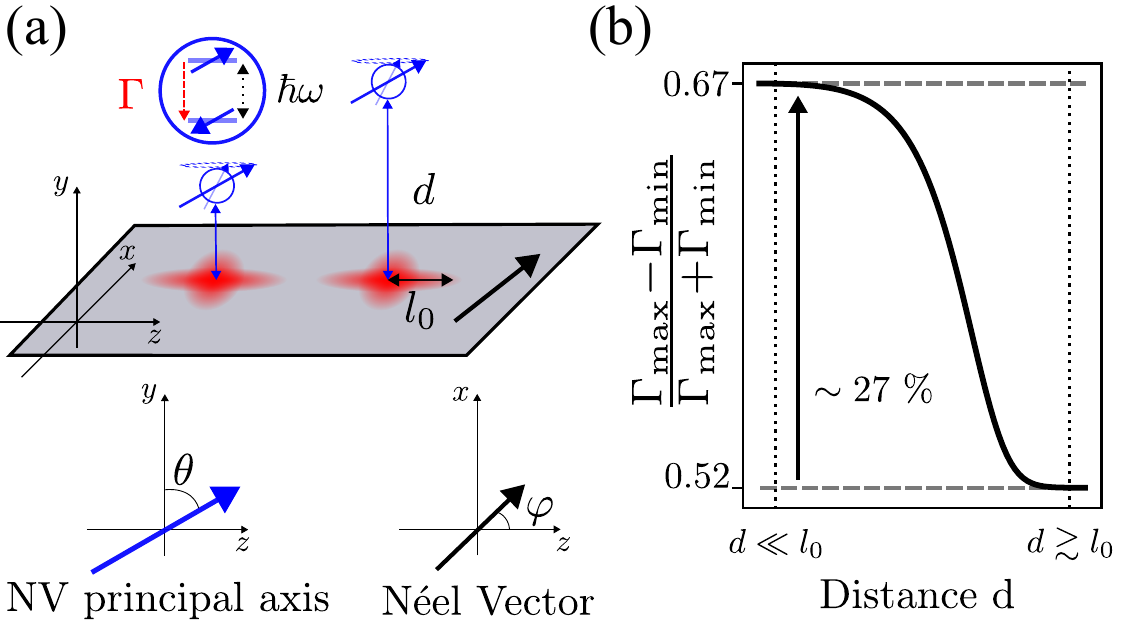}
  \caption{(a) Relaxometry of spin diffusion in altermagnets. The relaxation rate $\Gamma$ of a quantum impurity with transition frequency $\omega$ is monitored for different relative orientations, described by the angles $(\theta, \varphi)$, and at different distances $d$ to the altermagnet. The spin diffusion length $l_0$ sets the relevant distance scale. (b) The relaxation contrast over a set of relative orientations exhibits a non-trivial dependence with the distance, which is a fingerprint unique to altermagnetism. Its value can increase up to $\sim 27$ \%.}
\label{Fig:0}
\end{figure}

In this Letter, we propose a non-invasive sensing protocol based on the relaxometry properties of quantum spin impurities, such as nitrogen-vacancy (NV) centers in diamond~\cite{casola2018probing,hong2013nanoscale,kolkowitz2012sensing,dovzhenko2018magnetostatic,hsieh2019imaging,rovny2024new}, in order to distinguish altermagnets (ALMs) from conventional magnetic materials. Specifically, the dependence of the relaxation rate with the relative orientation between the NV principal axis and the N\'{e}el vector of the material changes with the distance between the NV and the sample -- a feature not present  in conventional magnets \cite{du2017controlandlocalmeasurement,wang_2022_noninvasive_measurements,Heitzer_2024_Characterization}. The spin diffusion coefficients of ALM insulators can then be obtained by analyzing the decay of the impurity's relaxation rate at different orientations with the distance from the sample material, as schematized in Fig.~\ref{Fig:0}(a). We propose a contrast function, exemplified in Fig.~\ref{Fig:0}(b), that depends exclusively on the diffusion coefficients, and exhibits an excursion of up to 27 \% with the NV-sample distance, for commercially available NVs, whilte it would be completely flat for conventional magnets, such as AFMs and ferromagnets. Our scheme is a feasible and non-invasive tool for probing spin diffusion in ALM insulators, a task so-far not accomplished in experiments, and which is valuable for devising new applications in which spin transport is pivotal, such as spin-based information processing.

\textit{Quantum impurity relaxometry  --} We consider the setup of Fig.~\ref{Fig:0}. A quantum impurity (QI), such as a NV center, is placed in the proximity of a two-dimensional magnetic material. In the following, we will keep our derivation general, and specialize to NVs whenever necessary. The relaxation rate $T_1$ of the impurity is then composed by an intrinsic contribution (e.g. due to decay into phonons for NVs), and a contribution due to the noisy stray magnetic  field of the sample. The QI magnetic moment, associated to a transition between two spin levels, couples to the stray magnetic field generated by the magnetic excitations of the material via dipole coupling. The noisy stray magnetic field causes transitions between the two levels with relaxation rate~\cite{flebus_2018_quantum_impurity_relaxometry, chatterjee2019diagnosingphases}
\begin{equation}\label{eq:rate}
\Gamma[\omega] = \gamma^2 \int dt e^{- i \omega t} 
\langle \{ \hat{B}^{(+)}_{\rm{QI}}(t),\hat{B}^{(-)}_{\rm{QI}}(0) \} \rangle/2, 
\end{equation}
where $\{\cdot, \cdot\}$ is the anticommutator, $\hat{B}^{(+)}_{\rm{QI}}(t) = \hat{B}_{x, \rm{QI}} + \hat{B}_{y, \rm{QI}}$ is the magnetic field operator associated with the spin fluctuations of the material, $\hbar \omega$ the energy  level splitting  of the quantum impurity and $\gamma$ the gyromagnetic ratio of the dipole transition. By measuring $T_1$, it is then possible to infer the additional decay of the NV due to the magnetic noise. 

The stray magnetic field correlations carry information about single and multiple magnon scattering processes in the sensed magnet \cite{langsjoen2012qubit,Agarwal_currentNoise2017,flebus_2018_quantum_impurity_relaxometry,Rodriguez-Nieva_1D_2018,chatterjee2019diagnosingphases,rodriguez2022probing,Heitzer_2024_Characterization}. \corr{The relaxation rate, thus depends on both transversal and longitudinal spin correlations \cite{flebus_2018_quantum_impurity_relaxometry,Heitzer_2024_Characterization}. The transversal correlations are associated with single magnon processes, which are relevant if the frequency, $\omega$, of the NV  is at resonance with the magnon frequency (plus a broadening due to magnon relaxation). For altermagnets, the typical magnon gap is $\sim$ THz \cite{smejkal_2023_chiral_magnons}, while the NV frequency is $\sim$ GHz, and thus contributions due to transverse spin correlations to the NV relaxation are negligible.} In this case, the main contribution to $\Gamma$ is the longitudinal correlation function $\langle \{ \hat{s}_\parallel(\vec{r},t),\hat{s}_\parallel(\vec{r}^\prime,0) \} \rangle$, where $\parallel$ refers to the component of the spin density operator $\hat{\vec{s}}$ parallel to the N\'{e}el vector. This correlation function is associated with two magnon processes and, thus, with spin transport. These conditions hold for a \corr{NV center} with $\omega \sim$ GHz \cite{mzyk2022relaxometry}, and an ALM in a weak bias field and with easy-axis anisotropy \cite{smejkal_2022_beyond_conventional,smejkal_2022_emerging_research,smejkal_2023_chiral_magnons,eto2025spontaneousmagnon}.

Using the fluctuation-dissipation theorem \cite{kubo_1966_fluctuation_dissipation}, we  can rewrite the relaxation rate~\eqref{eq:rate}  in terms of the imaginary part of the spin susceptibility $\chi^{\prime \prime}_{\parallel}(\omega, \vec{k})$~\cite{Agarwal_currentNoise2017,chatterjee2019diagnosingphases}:   
\begin{equation}
\label{relax:fin}
\Gamma[\omega] =  \frac{\hbar \gamma^2 \tilde{\gamma}^ 2}{2}{\rm{coth}} \left( \frac{ \hbar \beta  \omega}{2} \right) \int \frac{d^2 \vec{k}}{(2 \pi)^2}  \mathcal{C}_{\theta,\corr{\varphi}}(d,\vec{k}) \chi^{\prime \prime}_{\parallel}(\omega, \vec{k}),
\end{equation}
where $\tilde{\gamma}$ is the magnet's gyromagnetic ratio,$d$ is the QI-sample distance, \corr{and $\beta = 1/k_B T$, where $T$ is the temperature and $k_B$ is the Boltzmann constant}. The geometric factor $\mathcal{C}_{\theta,\corr{\varphi}} (d,\vec{k}) = (2 \pi)^2 k^2 e^{- 2 k d} \mathcal{F}_{\theta,\corr{\varphi}} (\phi_k)$ (with ${\rm{tan}}(\phi_k) = k_x/k_z$) is obtained from the magnetostatic Green's function of a dipole~\cite{guslienki2011magnetostaticgreensfunction}, {  which is commonly encountered in NV-based noise spectroscopy~\cite{Agarwal_currentNoise2017,chatterjee2019diagnosingphases,rodriguez2022probing},} and depends on the relative orientation between the NV principal axis (assumed to be in the $yz$ plane) and the in-plane N\'{e}el vector of the ALM via the function $\mathcal{F}_{\theta,\corr{\varphi}}(\phi_k)$, where the angles $(\theta, \corr{\varphi})$ describe the orientation of the NV principal axis and the N\'{e}el vector, as illustrated in  Fig. \ref{Fig:0}. We remark that our formalism would  also be suitable for N\'{e}el vectors with out-of-plane components. The coefficient $k^2 e^{- 2 k d}$ in $\mathcal{C}_{\theta,\corr{\varphi}}(d,\vec{k})$\corr{, which is a consequence of the dipolar nature of the interaction between the NV spin and the magnetic excitations,} introduces a momentum filter function: for $k \gtrsim 1/d$  the integrand is suppressed by the exponential factor $e^{- 2 k d}$, and at the same time it dampens it for $k\lesssim 1/d$ due to the term proportional to $k^2$. This means that we can single out the response function Eq.~\eqref{relax:fin}  around $k \simeq 1/d \equiv k_d$ in the momentum integral involved in the calculation of the relaxation rate~\eqref{relax:fin}, or in physical terms, only magnons with wavelengths $\simeq 1/d$ contribute. 
The central result of this Letter is that the relaxation rate in Eq.~\eqref{relax:fin} can distinguish between spin diffusion in ALMs and in more conventional magnets such as antiferromagnets (AFMs), as a result of the anisotropy of the diffusion response function $\chi_{\parallel}(\omega, \vec{k})$ in $k$-space -- a consequence of the alternating magnon band structure of an ALM \cite{smejkal_2022_beyond_conventional,smejkal_2023_chiral_magnons}.\\

\textit{Spin diffusion in ALMs --} We obtain the response function $\chi_{\parallel}$ in the diffusive limit following the phenomenological transport approach in~\cite{rezende_2016_bulk_magnon,rezende_2019_introduction} adapted to ALMs \cite{SM}. We assume that the magnon excess density $\delta n_{\alpha,\beta}$ (for $\alpha$ and $\beta$ the two magnon species) follows a diffusion equation, with a diffusion tensor $\overleftrightarrow{D}_{\alpha,\beta}$ that depends on the magnon band structure, and the magnon momentum relaxation rate \cite{eto2025spontaneousmagnon,SM}.

\begin{figure}
\centering
  \includegraphics[width=0.8 \columnwidth]{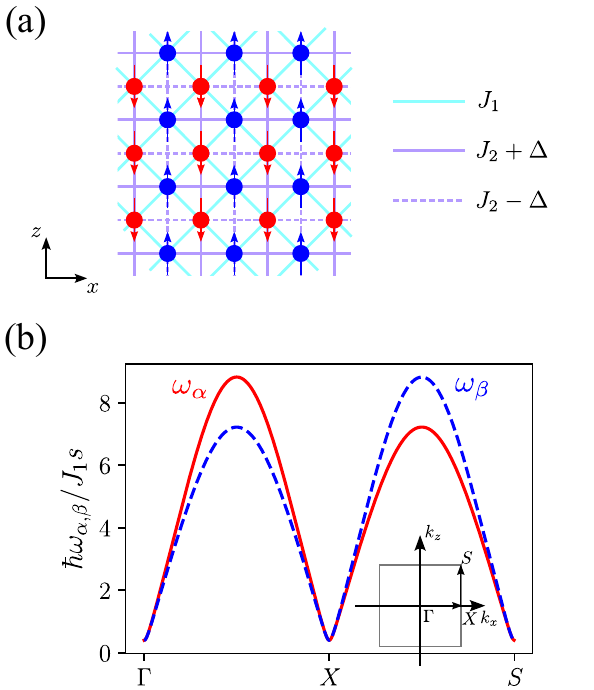}
  \caption{(a) Lieb lattice altermagnet. The different exchange constant are: $J_1$, the nearest neighbor exchange, and $J_2 \pm \Delta$ the next-nearest neighbor exchanges. The system is assumed in an in-plane N\'{e}el state due to an in-plane easy-axis anisotropy $J_{\rm{EA}}$. (b) Magnon frequencies $\omega_{\alpha,\beta}$ along the indicated path in the Brillouin zone for $J_2 = 0.5 J_1$, $\Delta = 0.1 J_1$, and $J_{\rm{EA}} = 0.01 J_1$.}
\label{Fig:01}
\end{figure}

As a simple way to model magnon bands in a 2D ALM \cite{PhysRevB.108.224421,cui_2023_efficient_spin,das_2024_realizing_altermagnetism,D4NR04053H,Lanzini2025Dual}, we resort to the Lieb lattice of Fig.~\ref{Fig:01}(a) with the Heisenberg Hamiltonian
\begin{equation}
\label{eq:HeisHam}
\begin{aligned}
\hat{\mathcal{H}} &=  J_1 \sum_{\langle \vec{r},\vec{r}^\prime \rangle} \hat{\vec{S}}_i \cdot \hat{\vec{S}}_j + \sum_{\langle \langle \vec{r},\vec{r}^\prime \rangle \rangle_\pm} (J_2\pm \Delta) \hat{\vec{S}}_i \cdot \hat{\vec{S}}_j \\&+ \frac{J_{\rm{EA}}}{2} \sum_{\vec{r}} \left(\hat{S}_{z,j} \right)^2,
\end{aligned}
\end{equation}
where $J_1$ is the nearest-neighbor exchange between sublattices with opposite spins, $(J_2\pm \Delta)$ are the next-nearest-neighbor exchange, which depend on the direction, and $J_{\rm{EA}}$ is the on site easy-axis anisotropy. We then perform a Holstein-Primakoff approximation to first order \cite{rezende_2019_introduction,SM}, and obtain a quadratic bosonic Hamiltonian describing the excitation on top of the N\'{e}el order. A standard diagonalization procedure yield the magnon frequencies $\omega_{\alpha,\beta}$ shown in Fig.~\ref{Fig:01}(b) around the indicated path in the Brillouin zone. These magnon bands display the alternating band splitting which, in this model, is parametrized by $\Delta$. In order to measure spin diffusion, the QI transition frequency $\omega$ has to lie within the magnon gap $\Delta_{\rm{mag}}=4s(16 J_1 J_{\rm{EA}} + J_{\rm{EA}}^2)^{1/2}/\hbar$, rendering any single-magnon process (e.g. annihilation of a magnon followed by a transition in the QI) off-resonant. This guarantees that longitudinal spin-correlations constitute the major contribution for Eq.~\eqref{eq:rate}. For a QI with $\omega \sim$ GHz and an ALM with exchange constant $J_1 \sim 10$ meV, even a weak anisotropy $J_{\rm{EA}}$ would satisfy such requirement.

The spin model considered here illustrates a d-wave ALM, with the magnetic lattice exhibiting a 4-fold rotation symmetry \cite{smejkal_2022_emerging_research}, which is also present in the magnonic band structure \cite{corticelli2022spinspacegroups}. In turn, we can show that the magnon diffusion tensor has the general form $D_{\alpha, xx} = D_{\beta, xx} = D_{\alpha, zz} =D_{\beta, zz} = D_1$, while  $D_{\alpha, xz} =-D_{\beta, xz}= D_2$ \cite{SM}. Such relations should hold for any d-wave ALM, and are a direct consequence of the symmetries of the Heisenberg Hamiltonian. We furthermore notice that for $\Delta =0$, i.e., in the absence of the ALM band splitting, $D_2 = 0$, in which case the model describes a Liebe lattice AFM. The isotropic diffusion of tensor of such an ALM is a consequence of the Heisenberg Hamiltonian -- the exchange constants are the same for all components of the spins, and are isotropic.

The spin diffusion response function is then obtained from the set of coupled linear diffusion equations for  $\mathcal{S}_\parallel = \hbar(- \delta n_{\alpha} + \delta n_{\beta})$, related to the longitudinal spin density, and $\mathcal{S}_\perp= \hbar(\delta n_{\alpha} + \delta n_{\beta})$, related to the total magnon density \cite{SM}. These equations can be solved in Fourier domain through $\mathcal{S}_\parallel(\omega, \vec{k}) = \chi_{\parallel}(\omega, \vec{k}) h$, with the response function given by \cite{SM}
\begin{equation}
\label{eq:fullrespfunc}
\chi_{\parallel}(\omega, \vec{k}) = \frac{\hbar \chi_0 \left[ D_1 k^2 + \frac{1}{\tau_s} - \frac{D_2^2 k^4 \cos^2(2 \phi_k)}{- i \omega +D_1 k^2 + 1/\tau_s} \right]}{- i \omega + D_1 k^2 + \frac{1}{\tau_s} - \frac{D_2^2 k^4 \cos^2(2 \phi_k)}{- i \omega +D_1 k^2 + 1/\tau_s}},
\end{equation}
where $\tau_s$ is the spin relaxation time. The dependence of $\chi_{\parallel}$ on the directionality in $k$-space due to factor $\propto k^4 \cos^2(2 \phi_k)$ is a unique trait of ALMs, not present in usual magnets \cite{flebus_2018_quantum_impurity_relaxometry,wang_2022_noninvasive_measurements,Heitzer_2024_Characterization}.

\textit{Relaxation rate and contrast -- }The response function $\chi_\parallel$ in Eq.~\eqref{eq:fullrespfunc} is anisotropic in $k$-space. As a consequence, the dependence of the relaxation rate on the relative orientation $(\theta,\varphi)$ at a given distance $d$ is different for ALMs when compared with conventional magnets. To better highlight this feature, we find it useful to define the relaxation contrast
\begin{equation}
\label{eq:contrast}
\mathcal{C}[d] = \frac{\Gamma_{\rm{max}}[d] -\Gamma_{\rm{min}}[d]}{\Gamma_{\rm{max}}[d] +\Gamma_{\rm{min}}[d]},
\end{equation}
which depends on the maximum and the minimum of the relaxation rate over a set of relative orientations at a given QI-sample distance. For an AFM, the spin diffusion susceptibility is isotropic, and, as a result of that Eq.~\eqref{eq:contrast} will be independent of the distance $d$, while  ALM exhibit a non-trivial dependence of the contrast with the QI-sample distance. To further analyze the contrast and the QI relaxation, we take the DC limit ($\omega \to 0$)\footnote{This approximation requires that $\omega \ll D_1 l_0/d^2$. We nevertheless emphasize that the plots shown in the paper are for a finite $\omega = 1/\tau_s$.}, which should be valid for a NV operating in the GHz regime.

The contrast is maximized when the QI axis is aligned with the magnet's plane, i.e. $\theta= \pi/2$, cf. Fig.~\ref{Fig:0}. We then consider variations in $\varphi$. The maximum relaxation rate occurs for $\varphi = 0$, while the minimum is at $\varphi = \pi/2$. The contrast for $d \gg \sqrt{D_1 \tau_s}$ is then given by $\mathcal{C}(d)\vert_{d \gg l_0} = 11/21 \approx 0.52$, which is independent of any microscopic parameter. In fact, the relaxation rate in this limit is  
\begin{equation}
\label{eq:far}
\begin{aligned}
\Gamma[\omega]\vert_{d \gg l_0} \propto \frac{l_0^4}{d^4} \int_0^{2 \pi} d  \phi_k \mathcal{F}_{\theta,\varphi} (\phi_k),
\end{aligned}
\end{equation}
where we have defined $l_0 = \sqrt{D_1 \tau_s}$, the isotropic spin diffusion length. Such a relaxation rate has exactly the same form one would get from a magnet with isotropic magnon bands \cite{flebus_2018_quantum_impurity_relaxometry,wang_2022_noninvasive_measurements}.
The value of the contrast is independent on whether the material is an ALM or an AFM, a consequence of the anisotropy of the magnetic noise driving the QI with respect to the orientation angle $\varphi$. The N\'{e}el vector sets the direction of the longitudinal modes probed by relaxometry, rendering the relaxation rate intrinsically anisotropic. ALMs induces further anisotropy contributions due to the magnon splitted bands, which, according to Eq.~\eqref{eq:fullrespfunc}, is washed away for wave-vectors $k \ll 1/l_0$. Thus, the NV is insensitive to such additonal trait intrinsic to ALMs when placed at distances $d \gg l_0$.

\begin{figure}
\centering
  \includegraphics[width=0.8 \columnwidth]{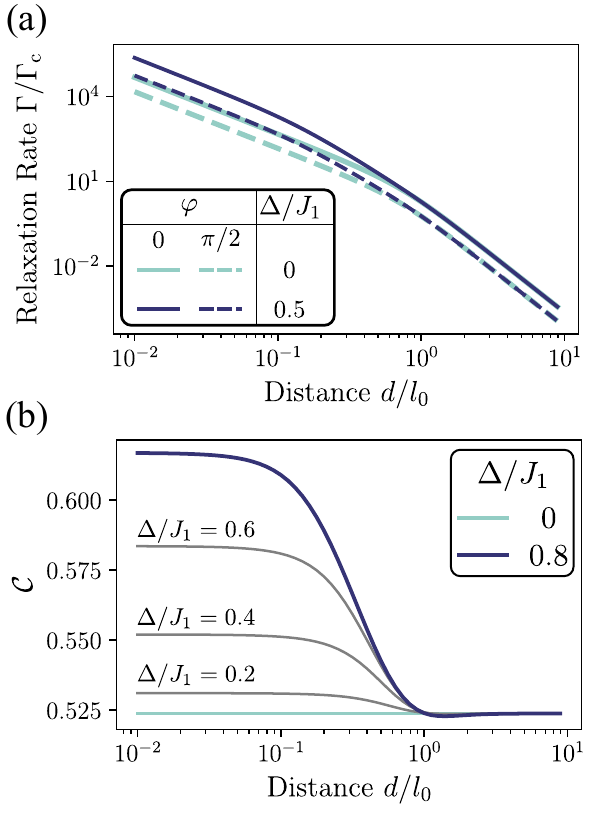}
  \caption{(a) QI relaxation rate as a function of the distance to the sample $d$ in units $\Gamma_{\rm{c}}$, given in the text, for different relative orientations. (b) Relaxation contrast  defined in Eq.~\eqref{eq:contrast} as a function of the distance $d$ for different values of the parameter $\Delta$. Parameters are: $J_2 = -0.8 J_1$, $J_{\rm{EA}} = 0.01 J_1$, and $\omega \tau_s = 1$.}
\label{Fig:02}
\end{figure}

At close distances $d \ll l_0$, the QI becomes sensitive to the intrinsic anisotropy of spin diffusion in ALMs, since the relaxation rate has a different dependence on the orientation angles. In this limit the relaxation rate is given by \footnote{The singularity of the integrand at $\phi_k = 0 , \pi/2, ...$ and $D_2 = D_1$ is an artifact of the approximation $\omega \sim 0$ that was taken to write Eq.~\eqref{eq:close}. The frequency $\omega$ is never zero, and thus the integrand is always finite. All plots shown in the paper are generated with the full expression including a non-vanishing frequency, which is shown in the SI.} 
\begin{equation}
\label{eq:close}
\begin{aligned}
\Gamma[\omega]\vert_{d \ll l_0} \propto \frac{D_1^2 l_0^2}{d^2}\int_0^{2 \pi} d  \phi_k \frac{\mathcal{F}_{\theta,\varphi} (\phi_k)}{D_1^2- D_2^2 \cos^2(2 \phi_k)}.
\end{aligned}
\end{equation}
While the relaxation rate exhibits the same scaling $d^{-2}$ as AFMs, the anisotropy of the response function Eq.~\eqref{eq:fullrespfunc} induces an additional dependence of the integrand from directionality in $k$-space in the above equation. In turn, the contrast in this limit is given by
\begin{equation}
\mathcal{C}(d\ll l_0) = \frac{5 D_1(D_1- \sqrt{D_1^2 -  D_2^2}) + 3 D_2^2}{3 D_1(D_1 - \sqrt{D_1^2-  D_2^2})+9 D_2^2}.
\end{equation}
In the case of an AFM, i.e. in the limit $D_2 \rightarrow 0$ with $D_1$ finite, the contrast yields $11/21$, which is the same value it has at  distances $d\gg l_0$. For an ALM, the contrast behaves non-trivially with distance,   exhibiting a larger value at closer distances. For $D_2 \approx D_1$ (its maximum value), the contrast at closer distances is $8/12 \approx 0.67$, which is an increase of $27$\% compared to the large distance limit, as shown in Fig.~\ref{Fig:0}(b). In general, the contrast depends on the ratios $D_2/D_1$ and $d/l_0$, and on the product $\omega \tau_s$ \cite{SM}, thus its behavior with respect to the other parameters can also be exploited for extracting intrinsic diffusion parameters. Such a behavior of the contrast is a consequence of the anisotropy of the response function in Eq.~\eqref{eq:fullrespfunc}, generated by the off-diagonal component of the diffusion tensor $D_2$.
We show in Fig.~\ref{Fig:02} the relaxation rate $\Gamma$, in units of the characteristic rate $\Gamma_c = \hbar \chi_0 \gamma^2 \tilde{\gamma}^2 k_b T \tau_s/l_0^4$, and the contrast $\mathcal{C}$ in as a function of the distance $d$ for different values of the parameter $\Delta$. The relaxation rates for ALMs exhibit a behavior with the distance $d$ similar to AFMs, as show in Fig.~\ref{Fig:02}(a). Nevertheless, the contrast, depicted in Fig.~\ref{Fig:02}(b), increases as the distance $d$ is decreased. This is a trait that is unique to ALMs, and a consequence of the anisotropic term of Eq.~\eqref{eq:fullrespfunc}.

\textit{Feasibility --} The first requirement of our sensing scheme is a QI at a distance $d \ll l_0$ from the two-dimensional ALM sample. The diffusion length can be estimate by $l_0 = s J_1 a \sqrt{\tau \tau_s}/\hbar$ (where $\tau$ is the magnon momentum relaxation time), where $a$ is the lattice constant, and $\tau$ is the magnon momentum relaxation. For the representative parameters $\tau_s \sim 10$ ns, $J_1 = 10$ meV, $a \sim 3$ $\AA$ and $\tau \sim 10$ ps, we obtain $l_0 \approx 2.0$ $\mu$m, well within commercially available all-diamond probes which can achieve distances of $\sim 50$ nm \cite{Finco_2023_single_spin}. 

The relaxation rate induced by the spin transport should dominate the intrinsic relaxation rate $\Gamma^ {(0)}$ experienced by the QI in the absence of magnetic noise. This is particularly relevant for the measurement of the spin diffusion noise at farther distances, for which the spin-diffusion induced relaxation is weaker. \corr{$\Gamma^ {(0)}$ includes contributions due to coupling to phonons \cite{Gugler_2018Abinitio,Norambuena2018spinlattice}, magnetic noise \cite{Gugler_2018Abinitio,Norambuena2018spinlattice}, and due to electric and strain noise \cite{Sangtawesin2019Origins,candido2024interplay}. The first two sources can be minimized by operating the system at low temperatures and using high-quality diamonds \cite{Irber2021RobustallOptical}. The electric and strain contributions depend mainly on the distance between the NV center and the surface of the diamond \cite{Sangtawesin2019Origins,candido2024interplay}, and can be optimized with proper fabrication techniques.}

The variation in contrast with the distance is already significant around $d \sim l_0$, as shown in Fig.~\ref{Fig:02}(b). The characteristic relaxation rate $\Gamma_c$, for the above representative parameters, is $\Gamma_c \sim 25$ Hz \cite{SM}, which is comparable to $\Gamma^ {(0)}$ at 200 K \cite{cambria_2023_temperature_dependent}. We take the results of Fig.~\ref{Fig:02} as a benchmark for measurement requirements: at the far distance $d \sim \, 1.6 \, \mu$m, the measurement of the $0.53$ contrast would require the distinction between a total relaxation rate $102$ Hz and $31$ Hz. At closer distances $d < 0.1 l_0$, we obtain relaxation rates of few hundreds of kHz. The measurement of those values are within the capabilities of state-of-the-art systems \cite{cambria_2023_temperature_dependent}, \corr{but would require the system to be operated at temperatures below 200K. The} maximum contrast enhancement, from $d \sim l_0$ to $d \ll l_0$, is $\sim 17$ \%. \corr{Measurements with different axis orientations can be taken by rotating the diamond while keeping the sample fixed \cite{Weggler2020Determination} or using two NVs with different axis orientations hosted in the same diamond. The NV axis can be set by fabrication \cite{Edmonds2012Production,Klink2025Fabrication}, with the maximum relative angle between two QIs being $\pi/3$, which would yield a reduction of the maximum contrast by $\sim 0.8$ compared with the ideal case.} \corr{The optical readout of NV spins also has technical limitations, but is an established technique which new developments that demonstrate high-fidelity readout capabilities \cite{Irber2021RobustallOptical}. Typical measured values of $T_1$ exhibit errors $<$ 10 $\%$ \cite{cambria_2023_temperature_dependent,Barbosa2024Temperature}, and could allow the measurements of ALMs with a moderate ratio $D_2/D_1$.}

\textit{Perspectives -} In this Letter, we discussed the quantum noise spectroscopy of insulating ALMs using NV centers, where the carriers of spin in the material are localized. In a similar way, quantum noise spectroscopy can also be used to study the properties of metallic ALMs (MALMs). In this case, additional considerations arise. Similar to the noise spectroscopy of generic metallic systems, one must account for the Johnson noise due to current fluctuations~\cite{langsjoen2012qubit,Agarwal_currentNoise2017,Chatterjee_2Dsupercond2022}. The notable feature of MALMs, however, is the mobile nature of magnetic moments in conjunction with strong spin-orbit interaction, leading to a qualitatively distinct dynamical spin response function and the resulting magnetic noise. The presence of spin-momentum locking leads to a rich dependence of the relaxation on the orientation of the NV center. Moreover, nodal structure of altermagneto electronic structure~\cite{smejkal_2022_emerging_research}, significantly alters the magnetic noise spectrum. 
The gapless nature of the magnetic excitations also makes $T_2$-based spectroscopy a particularly powerful tool for probing the spin dynamics in such systems, as discussed below.

An alternative approach to probing ALMs is through $T_2$ relaxometry~\cite{machado_2023_quantum_noise}. The typically stronger $T_2$ signal, as compared to $T_1$ noise, together with its operational low-frequency range,  makes it well-suited for studying systems with low-energy and gapless modes. These include, the longitudinal diffusive modes, hydrodynamic sound modes, as well as nodal magnons in MALMs mentioned above. Recent advances in NV-based sensing allow for multi-qubit noise spectroscopy, which can provide spatially resolved information on magnetic noise correlations~\cite{rovny2022nanoscale,rovny2024new}. This technique offers powerful means to probe non-invasively spatio-temporal spreading of correlations in ALMs~\cite{Hosseinabadi_twoNV2025}.

\textit{Acknowledgments -}  H.H. and J.M. are grateful to E. Demler, P. Dolgirev, S. Gopalakrishnan for related discussions on NV sensing. VASVB acknowledges financial support from the Contrat Triennal 2021-2023 Strasbourg Capitale Europeenne. JS and LS acknowledge support by Deutsche Forschungsgemeinschaft (DFG, German
Research Foundation) - DFG (Project 452301518) and TRR 288 – 422213477 (project A09 and A12). LS acknowledges support by the ERC Starting Grant No. 101165122. JM acknowledges support by the DFG through the grant HADEQUAM-MA7003/3-1.  We acknowledge support from the Dynamics and Topology Centre, funded by the State of Rhineland Palatinate.

%
%
\newpage

\onecolumngrid

\setcounter{equation}{0}
\setcounter{figure}{0}
\setcounter{table}{0}
\renewcommand{\theequation}{S\arabic{equation}}
\renewcommand{\thefigure}{\arabic{figure}}
\renewcommand{\figurename}{SUPPLEMENTARY FIGURE}
\renewcommand{\tablename}{SUPPLEMENTARY TABLE}

\section*{Supplementary Material: Quantum impurity sensing of altermagnets}

\section{Magnetostatic Green's function and the geometric factor of the relaxation rate}

In this section we present explicitly the relation between the stray magnetic field $\hat{\vec{B}}_{\rm{QI}}$ in Eq.(1) and the spin density $\hat{\vec{s}}$ and obtain explicitly the geometric factor $\mathcal{C}_{\{\Phi \}} (d, \vec{k})$ of Eq.~(2). Here, we indicate the set of orientation angles by $\{\Phi \} = (\theta, \varphi)$, where $\theta$ and $\varphi$ defined in the main text and in Fig.~1. The stray magnetic field $\hat{\vec{B}}_{\rm{QI}}$ is given by the the stray magnetic field $\hat{\vec{B}}$ written in the quantum impurity (QI) principal axis frame. We keep to the setup of Fig.~1, and consider that the QI principal axis lies in the yz plane, such that $\hat{\vec{B}}_{\rm{QI}} = \mathcal{R}_x(\theta) \cdot \hat{\vec{B}}$, where $\mathcal{R}_x(\theta)$ is a rotation around the $x$ axis. 

The field $\hat{\vec{B}}(\vec{r}, t)$ and the spin density $\vec{s}$ are related by the magnetostatic Green's function \cite{dovzhenko2018magnetostatic}:
\begin{equation}
\hat{\vec{B}}(\vec{r},t) = \tilde{\gamma} \int d^3 \vec{r}^\prime \overleftrightarrow{\mathcal{D}}(\vec{r}, \vec{r}^\prime) \cdot \hat{\vec{s}}(\vec{r}^\prime, t), 
\end{equation}
where $\overleftrightarrow{\mathcal{D}}(\vec{r}, \vec{r}^\prime) = - \nabla_{\vec{r}} \nabla_{\vec{r}^\prime} (1/\vert \vec{r} - \vec{r}^\prime \vert)$. The elements of the magnetostatic Green's function for a thin film read
\begin{equation}
\begin{aligned}
\mathcal{D}_{jl}(\vec{r},\vec{r}^\prime)  &= - \frac{1}{2 \pi} \int d^2 \vec{k} \frac{k_j k_l}{k} e^{-k \vert y - y^\prime \vert} e^{i \vec{k} \cdot (\vec{\rho} -\vec{\rho}^\prime)} \quad \quad (j,l= x,z), \\
\mathcal{D}_{yj}(\vec{r},\vec{r}^\prime)  &= - \frac{i \rm{sign}(y - y^\prime)}{2 \pi} \int d^2 \vec{k} k_j e^{-k \vert y - y^\prime \vert} e^{i \vec{k} \cdot (\vec{\rho} -\vec{\rho}^\prime)} \quad \quad (j= x,z), \\
\mathcal{D}_{yy}(\vec{r},\vec{r}^\prime)  &= - \frac{1}{2 \pi} \int d^2 \vec{k} k e^{-k \vert y - y^\prime \vert} e^{i \vec{k} \cdot (\vec{\rho} -\vec{\rho}^\prime)},
\end{aligned}
\end{equation}
where $k = \sqrt{k_x^2 + k_z^2}$ and $\vec{\rho} = x \vec{e}_x + z \vec{e}_z$ (with a similar definition for $\vec{\rho}^\prime$). Furthermore, the spin density $\hat{\vec{s}}(\vec{r}^\prime, t)$ is proportional to $\delta(y)$, and can be written as $\hat{\vec{s}}(\vec{r}, t) = \mathcal{R}_{y}(\varphi)\hat{\vec{s}}_{N}(\vec{r}, t)$, where $\hat{\vec{s}}_{N}$ is the spin density written in a frame of the N\'{e}el vector, and $\mathcal{R}_{y}(\varphi)$ is a rotation around the $y$ axis. We can then write $\hat{B}_{{\rm{QI}} }^{(-)} = \hat{B}_{{\rm{QI}} , x} - i\hat{B}_{{\rm{QI}}, y}$, which is required for calculating the relaxation rate, as
\begin{equation}
\label{eq:BQIminus}
\begin{aligned}
\hat{B}_{{\rm{QI}}}^{(-)}(d,t) &= \tilde{\gamma} \int d^2 \vec{\rho}^\prime \Big[ \mathcal{A}^{-}_{\theta,\varphi}(\vec{\rho}^\prime,d) \hat{s}^{(-)}(\vec{\rho}^\prime, t) +\mathcal{B}^{-}_{\theta,\varphi}(\vec{\rho}^\prime,d) \hat{s}^{(+)}(\vec{\rho}^\prime, t) +\mathcal{C}^{-}_{\theta,\varphi}(\vec{\rho}^\prime,d) \hat{s}_{\parallel}(\vec{\rho}^\prime, t) \Big],
\end{aligned}
\end{equation}
where the components of the spin operators are in the frame of the N\'{e}el vector. The relaxation rate in Eq.~(1) has contribution from different correlators. As argued in the main text, we the most relevant for the setup we envision are the ones proportional to $\langle \{\hat{s}_{\parallel}(\vec{\rho}^\prime, t), \hat{s}_{\parallel}(\vec{\rho}^{\prime, \prime}, 0)   \} \rangle$, where $\{\cdot, \cdot  \}$ denotes the anticommutator. We substitute Eq.~\eqref{eq:BQIminus} into Eq.~(1) and write the corresponding integrals in the Fourier domain keeping only the aforementioned spin correlators, which gives
\begin{equation}
\Gamma[\omega] = \gamma^2 \tilde{\gamma}^2 \int \frac{d^2 \vec{k}}{(2 \pi)^2} \mathcal{C}_{\{\Phi \}} (d,\vec{k}) C_{\parallel}(\vec{k},\omega),
\label{GammaInit}
\end{equation}
where $C_{\parallel}(\vec{k},\omega)$ is the Fourier transform of the spin correlator. $\mathcal{C}_{\{\Phi \}} (d,\vec{k})$ can be obtained via the factor $\mathcal{C}^{-}_{\theta, \varphi}$ of Eq:~\eqref{eq:BQIminus}, which uses the Fourier domain form of the elements of the magnetostatic Green's tensor. After some algebra we have
\begin{equation}
\label{eq:OrienFact}
\begin{aligned}
\mathcal{C}_{\{\Phi \}} (d,\vec{k}) &= (2 \pi)^2 k^2 e^{- 2 k d} \mathcal{F}_{\{ \Phi \}} (\phi_k) \\
\mathcal{F}_{\{ \Phi \}} (\phi_k) &= \cos ^2(\varphi ) \cos ^2\left(\phi _k\right)
   \left(\cos ^2(\theta ) \sin ^2\left(\phi
   _k\right)+\left(\sin (\theta )+ \cos
   \left(\phi _k\right)\right){}^2\right)
   \\
   &\quad+\sin
   ^2\left(\phi _k\right) \left(\cos ^2(\theta
   ) \sin ^2(\varphi ) \sin ^2\left(\phi
   _k\right)+\cos ^2(\theta ) \sin (2 \varphi
   ) \sin \left(\phi _k\right) \cos \left(\phi
   _k\right)+\sin ^2(\varphi ) \left(\cos
   \left(\phi _k\right)-\sin (\theta
   )\right){}^2\right) \\
   &\quad+2 \sin (\varphi ) \cos
   (\varphi ) \sin \left(\phi _k\right) \cos
   \left(\phi _k\right) \left(\cos \left(\phi
   _k\right)-\sin (\theta )\right) \left(\sin
   (\theta )+2 \cos \left(\phi
   _k\right)\right)
\end{aligned}
\end{equation}
The final form of the relaxation rate appearing in Eq.~(2) is then obtained by relating $C_{\parallel}(\vec{k},\omega)$ to the imaginary part of the response function $\chi_\parallel$ via the fluctuation-dissipation theorem \cite{kubo_1966_fluctuation_dissipation}.

\new{To connect our results with the literature, we consider now the case in which the Fourier transform of the spin correlator $C_\parallel (\vec{k},\omega)$, appearing in Eq.~\eqref{GammaInit}, is independent of the direction of the $k$-vector, i.e. $C_\parallel (\vec{k},\omega)=C_\parallel (k,\omega)$, where $k = \vert \vec{k} \vert$. In this case, the relaxation rate reads
\begin{equation}
\begin{aligned}
\Gamma[\omega] &= \gamma^2 \tilde{\gamma}^2 \int \frac{d^2 \vec{k}}{(2 \pi)^2} \mathcal{C}_{\{\Phi \}} (d,\vec{k}) C_{\parallel}(k,\omega) =(\gamma \tilde{\gamma})^2\int \frac{dk d\varphi_k}{(2 \pi)^2} k \, \mathcal{C}_{\{\Phi \}} (d,\vec{k}) C_{\parallel}(k,\omega).
\end{aligned}
\end{equation}
Using now Eq.~\eqref{eq:OrienFact} we have
\begin{equation}
\begin{aligned}
\Gamma[\omega] &= (\gamma \tilde{\gamma})^2\int dk d\phi_k k^3   e^{- 2 k d} \mathcal{F}_{\{ \Phi \}} (\phi_k) C_{\parallel}(k,\omega) \\ &=(\gamma \tilde{\gamma})^2\left[\int d\phi_k \mathcal{F}_{\{ \Phi \}} (\phi_k) \right] \left[\int dk k^3   e^{- 2 k d} C_{\parallel}(k,\omega)\right] \\
&=f(\{ \Phi\}) \left[\int dk k^3   e^{- 2 k d} C_{\parallel}(k,\omega)\right],
\end{aligned}
\end{equation}
where
\begin{equation}
f(\{ \Phi\}) = (\gamma \tilde{\gamma})^2\left[\int d\phi_k \mathcal{F}_{\{ \Phi \}} (\phi_k) \right].
\end{equation}
Although being a rather long calculation, the angular integral involves only trigonometric relations. Performing it gives us
\begin{equation}
\label{eq:ours}
f(\{ \Phi\}) = \frac{(\gamma \tilde{\gamma})^2\pi}{4}(5 - \cos^2 \theta + 2 \cos^2 \varphi  - 2  \cos^2 \varphi \cos^2 \theta).
\end{equation}
This is exactly the same geometric factor shown in the SI of Ref.~\cite{wang_2022_noninvasive_measurements} (Eq.~S16, after simple algebraic manipulations) for $\vartheta = 0$, which corresponds to in-plane N\'{e}el vector, as is the case considered here.}

\section{Magnon diffusion tensor}

We present here the full formula for the magnon diffusion tensor $\overleftrightarrow{D}$. We follow the procedure used for ferro- and antiferromagnets \cite{rezende_2016_bulk_magnon,rezende_2019_introduction}. Our starting point is the excess density $\delta n_{\xi} (\vec{r})$ of each magnon species $\xi = \alpha,\beta$ is
\begin{equation}
\delta n_{\xi} (\vec{r}) = \frac{1}{(2 \pi)^2} \int d^2 k \left( n_{\xi, \vec{k}}(\vec{r}) - n_{\xi, k}^0 \right),
\end{equation}
where $n_{\xi, k}^0$ is the equilibrium distribution of a given magnon mode (the Bose-Einstein distribution). We can also write the excess magnon current is 
\begin{equation}
\vec{j}_{\xi,\rm{Tot}} = \frac{1}{(2 \pi)^3} \int d^2 k \vec{v}_{\xi \vec{k}} \left[n_{\xi, \vec{k}}(\vec{r}) - n_{\xi, k}^0 \right],
\end{equation}
 $\vec{v}_{\xi \vec{k}} = \partial \omega_{\xi \vec{k}}/\partial \vec{k}$ is the magnon velocity. The steady-state excess distribution is given then by the solution of the Boltzmann equation
\begin{equation}
\label{eqd:01}
n_{\xi, \vec{k}}(\vec{r}) - n_{\xi, k}^0 = - \tau_{\xi, \vec{k}} \vec{v}_{\xi \vec{k}} \cdot \nabla n_{\xi, \vec{k}}(\vec{r}),
\end{equation}
where $\tau_{\xi, \vec{k}}$ is the momentum relaxation rate.

The excess magnon current density can then be decomposed in two components \cite{rezende_2016_bulk_magnon}: one related to changes on the equilibrium magnon distribution $n_{\xi, k}^0$ due to the change of some parameter, such as the temperature, and the contribution due to \textit{spatial accumulation of magnons}. To describe the magnon diffusive properties, we will focus on the latter which is given by
\begin{equation}
\label{eqd:02}
\vec{j}_{\xi} = - \frac{1}{(2 \pi)^2} \int d^2 k \tau_{\xi, \vec{k}} \vec{v}_{\xi \vec{k}} \left[\vec{v}_{\xi \vec{k}} \cdot \nabla (n_{\xi, \vec{k}}(\vec{r}) - n_{\xi, k}^0) \right].
\end{equation}
We can then write such a current in terms of a magnon diffusion tensor. For that, we consider the following form for the excess magnon distribution
\begin{equation}
\delta n_{\xi, \vec{k}}(\vec{r})=n_{\xi, \vec{k}}(\vec{r}) - n_{\xi, k}^0 = n_{\xi, k}^0 \varepsilon_{\xi,\vec{k}} g(\vec{r}),
\end{equation}
where $\varepsilon_{\xi,\vec{k}} = \hbar \omega_{\xi, \vec{k}}$, and thus the total magnon density accumulation is
\begin{equation}
\begin{aligned}
\delta n_{\xi} (\vec{r}) &= \frac{1}{(2 \pi)^2} \int d^2 k \left( n_{\xi, \vec{k}}(\vec{r}) - n_{\xi, k}^0 \right) \\
&= \frac{1}{(2 \pi)^3} \int d^2 k n_{\xi,\vec{k}}^0 \varepsilon_{\xi,\vec{k}} g(\vec{r}) = \frac{g(\vec{r})}{(2 \pi)^2} \int d^3 k n_{\xi, k}^0 \varepsilon_{\xi,\vec{k}} .
\end{aligned}
\end{equation}
We call from now on
\begin{equation}
I_0 = \frac{1}{(2 \pi)^2}\int d^2 k n_{\xi, \vec{k}}^0 \varepsilon_{\xi,\vec{k}},
\end{equation}
and thus
\begin{equation}
 g(\vec{r}) = \frac{\delta n_{\xi} (\vec{r})}{I_0}.
\end{equation}
We have then
\begin{equation}
\nabla \delta n_{\xi, \vec{k}}(\vec{r}) = n_{\xi, k}^0 \varepsilon_{\xi,\vec{k}} \nabla g(\vec{r}) =\frac{n_{\xi, k}^0 \varepsilon_{\xi,\vec{k}}}{I_0} \nabla \delta n_{\xi} (\vec{r}),
\end{equation}
and thus Eq.~\eqref{eqd:02} reads
\begin{equation}
\vec{j}_{\xi} = - \frac{1}{(2 \pi)^2} \int d^2 k \frac{n_{\xi, k}^0 \varepsilon_{\xi,\vec{k}}}{I_0} \tau_{\xi, \vec{k}} \vec{v}_{\xi \vec{k}} \left[\vec{v}_{\xi \vec{k}} \cdot  \nabla \delta n_{\xi} (\vec{r}) \right] =  - \frac{1}{I_0 (2 \pi)^2} \int d^2 k n_{\xi, k}^0 \varepsilon_{\xi,\vec{k}} \tau_{\xi, \vec{k}} \vec{v}_{\xi \vec{k}} \left[\vec{v}_{\xi \vec{k}} \cdot  \nabla \delta n_{\xi} (\vec{r}) \right].
\end{equation}
We write then for each component of the magnon excess current
\begin{equation}
j_{i, \xi} = - \left[\sum_{j } \frac{1}{I_0 (2 \pi)^2} \int d^2 k n_{\xi, k}^0 \varepsilon_{\xi,\vec{k}} \tau_{\xi, \vec{k}} v_{i,\xi \vec{k}} v_{j, \xi \vec{k}}    \right] \nabla_j \delta n_{\xi} (\vec{r}),
\end{equation}
from where we identify the magnon diffusion tensor element as
\begin{equation}
\label{Eq:DiffTensor}
D_{\xi, ij} =   \frac{1}{I_0 (2 \pi)^2} \int d^2 k n_{\xi, k}^0 \varepsilon_{\xi,\vec{k}} \tau_{\xi, \vec{k}} v_{i,\xi \vec{k}} v_{j, \xi \vec{k}}  ,
\end{equation}
and thus
\begin{equation}
j_{i, \xi} = - \sum_{j} D_{\xi, ij}   \nabla_j \delta n_{\xi} (\vec{r}).
\end{equation}

Given the above considerations, we can now write a diffusion equation for each magnon branch as
\begin{equation}
\label{eqs:DiffMag}
\begin{aligned}
\partial_t \delta n_{\alpha} +\nabla \cdot \vec{j}_{\alpha} &= -\frac{1}{\tau_s} (\delta n_{\alpha} + \chi_0 h), \\
\partial_t \delta n_{\beta} +\nabla \cdot \vec{j}_{\beta} &= -\frac{1}{\tau_s} (\delta n_{\beta} - \chi_0 h),
\end{aligned}
\end{equation}
where $h$ is a generalized force thermodynamically conjugated to $\delta n_{\xi}$, the static spin susceptibility is indicated by $\chi_0$, and the associated chemical potential is $(\delta n_{\xi}/\chi_0 \pm h)$ \cite{fang_2022_generalizedmodel}, with the different signs corresponding to the different magnon species \cite{flebus_2019_chemical_potential}. The spin relaxation time is indicated by $\tau_s$. The magnon currents are then given by
\begin{equation}
\begin{aligned}
\vec{j}_{\alpha} &= - \overleftrightarrow{D}_\alpha \cdot \nabla(\delta n_\alpha + \chi_0 h), \\
\vec{j}_{\beta} &= - \overleftrightarrow{D}_\beta \cdot \nabla(\delta n_\beta- \chi_0 h). \\
\end{aligned}
\end{equation}
For the equations, we have used the fact that, because each branch has a different spin angular momentum, the static susceptibility and the chemical potential need to have opposite signs \cite{flebus_2019_chemical_potential}.

\subsection{Magnon bands in the Lieb lattice altermagnet}

To calculate the elements of the diffusion tensor given in Eq.~\eqref{Eq:DiffTensor}, we need first the magnon velocities, which in turn depend on the magnon band structure. We consider a 2D altermagnet described by a Heisenberg Hamiltonian in a Lieb lattice, shown in Fig.~2 in the main text.
\begin{equation}
\label{eq:HeisHam}
\begin{aligned}
\hat{\mathcal{H}} &=  J_1 \sum_{\langle \vec{r},\vec{r}^\prime \rangle} \hat{\vec{S}}_i \cdot \hat{\vec{S}}_j + \sum_{\langle \langle \vec{r},\vec{r}^\prime \rangle \rangle_\pm} (J_2\pm \Delta) \hat{\vec{S}}_i \cdot \hat{\vec{S}}_j \\&+ \frac{J_{\rm{EA}}}{2} \sum_{\vec{r}} \left(\hat{S}_{z,j} \right)^2,
\end{aligned}
\end{equation}
where $J_1$ is the nearest-neighbor exchange between sublattices with opposite spins, $(J_2\pm \Delta)$ are the next-nearest-neighbor exchange, which depend on the direction, and $J_{\rm{EA}}$ is the on site easy-axis anisotropy. We first decompose the Hamiltonian in terms of intra- and intersublattice exchanges:
\begin{equation}
\label{eq:HeisHam2}
\begin{aligned}
\hat{\mathcal{H}} &=  J_1 \sum_{ \vec{r}_i, \{ \vec{\delta}_e \}} \hat{\vec{S}}_{A, \vec{r}_i} \cdot \hat{\vec{S}}_{B,\vec{r}_i + \vec{\delta}_e } + \frac{J_{\rm{EA}}}{2} \sum_{\vec{r}_i} \left[ \left(\hat{S}_{z, A, \vec{r}_i} \right)^2 + \left(\hat{S}_{z, B, \vec{r}_i} \right)^2 \right] \\
&+\sum_{\vec{r}_i, \{ \vec{\delta}_{1} \}}\left[  (J_2 + \Delta)
\hat{\vec{S}}_{A,\vec{r}_i} \cdot \hat{\vec{S}}_{A,\vec{r}_i + \vec{\delta}_{1}}  +  (J_2 - \Delta)\hat{\vec{S}}_{B,\vec{r}_i} \cdot \hat{\vec{S}}_{B,\vec{r}_i + \vec{\delta}_{1}} \right] \\
&+\sum_{\vec{r}_i, \{ \vec{\delta}_{2} \}}\left[  (J_2 - \Delta)
\hat{\vec{S}}_{A,\vec{r}_i} \cdot \hat{\vec{S}}_{A,\vec{r}_i + \vec{\delta}_{2}}  +  (J_2 + \Delta)\hat{\vec{S}}_{B,\vec{r}_i} \cdot \hat{\vec{S}}_{B,\vec{r}_i + \vec{\delta}_{2}} \right].
\end{aligned}
\end{equation}
The sets of vectors describing the positions of nearest and next-nearest neighbors are
\begin{equation}
\begin{aligned}
\{\vec{\delta}_e\}&: \quad \vec{\delta}_e^{(1)} = \frac{a}{2} \vec{e}_x +\frac{a}{2} \vec{e}_z, \quad \quad \vec{\delta}_e^{(2)} = -\vec{\delta}_e^{(1)},\\
&\quad \quad \vec{\delta}_e^{(3)} = \frac{a}{2} \vec{e}_x -\frac{a}{2} \vec{e}_z, \quad \quad \vec{\delta}_e^{(4)} = -\vec{\delta}_e^{(3)}. \\
\{\vec{\delta}_1\}&: \quad \vec{\delta}_1^{(1)} = a \vec{e}_x, \quad \quad \vec{\delta}_1^{(2)}= - \vec{\delta}_1^{(1)}. \\
\{\vec{\delta}_2\}&: \quad \vec{\delta}_2^{(1)} = a \vec{e}_z, \quad \quad \vec{\delta}_2^{(2)}= - \vec{\delta}_2^{(1)}.
\end{aligned}
\end{equation}

We then use the standard Holstein-Primakoff method, by first writting the spin operators in terms of the bosonic creation and annhiliation operator $\hat{a}_{\vec{r}_i}$ and $\hat{b}_{\vec{r}_i}$:
\begin{equation}
\begin{aligned}
\hat{S}_{A, \vec{r}_i}^{(+)} = \sqrt{2s} \hat{a}_{\vec{r}_i}, \hspace{0.5 cm} \hat{S}_{A, \vec{r}_i}^{(-)} &= \sqrt{2s} \hat{a}_{\vec{r}_i}^\dagger ,\hspace{0.5 cm} \hat{S}_{A, \vec{r}_i}^{(z)} = s- \hat{a}_{\vec{r}_i}^\dagger \hat{a}_{\vec{r}_i}, \\
\hat{S}_{B, \vec{r}_i}^{(+)} = \sqrt{2s} \hat{b}_{\vec{r}_i}^\dagger, \hspace{0.5 cm} \hat{S}_{B, \vec{r}_i}^{(-)} &= \sqrt{2s} \hat{b}_{\vec{r}_i}, \hspace{0.5 cm} \hat{S}_{B, \vec{r}_i}^{(z)} = -s+ \hat{b}_{\vec{r}_i}^\dagger \hat{b}_{\vec{r}_i},
\end{aligned}
\end{equation}
here $\hat{a}_{\vec{r}_i}$ and $\hat{b}_{\vec{r}_i}$ We substitute the above expressions into Eq.~\eqref{eq:HeisHam2} and Fourier transform the operators:
\begin{equation}
\hat{a}_{\vec{r}_i} = \frac{1}{\sqrt{N}} \sum_{\vec{k}} e^{i \vec{k} \cdot \vec{r}_i} \hat{a}_{\vec{k}},
\end{equation}
with a similar expression for $\hat{b}$. We get the following bosonic Hamiltonian
\begin{equation}
\label{eq:Ham}
\begin{aligned}
\frac{\hat{\mathcal{H}}}{\hbar}&= \sum_{\vec{k}} \left[\omega_a(\vec{k}) \hat{a}^\dagger_{\vec{k}} \hat{a}_{\vec{k}} + \omega_b(\vec{k}) \hat{b}^\dagger_{\vec{k}} \hat{b}_{\vec{k}}+ g_{ab}(\vec{k}) \left( \hat{a}_{\vec{k}} \hat{b}_{-\vec{k}} +\hat{a}^\dagger_{\vec{k}} \hat{b}^\dagger_{-\vec{k}} \right) \right],
\end{aligned}
\end{equation}
where
\begin{equation}
\label{freqs}
\begin{aligned}
\hbar \omega_a(\vec{k}) &= 4 s \left(J_1 - 2 J_2 + \frac{J_{\rm{EA}}}{2} \right) + 8 s J_2 \cos\left( \frac{a}{2} k_z +\frac{a}{2} k_x \right) \cos \left(\frac{a}{2} k_z -\frac{a}{2} k_x \right) +8 s \Delta \sin \left( \frac{a}{2} k_z +\frac{a}{2} k_x \right) \sin \left(\frac{a}{2} k_z -\frac{a}{2} k_x \right), \\
\hbar \omega_b(\vec{k}) &= 4 s \left(J_1 - 2 J_2 + \frac{J_{\rm{EA}}}{2} \right) + 8 s J_2 \cos \left(\frac{a}{2} k_z +\frac{a}{2} k_x \right) \cos \left(\frac{a}{2} k_z -\frac{a}{2} k_x \right) -8 s \Delta \sin \left(\frac{a}{2} k_z +\frac{a}{2} k_x \right) \sin \left(\frac{a}{2} k_z -\frac{a}{2} k_x \right), \\
\hbar g_{ab}(\vec{k})&= 2 s J_1 \left[ \cos\left(\frac{a}{2} k_z +\frac{a}{2} k_x \right) + \cos\left(\frac{a}{2} k_z -\frac{a}{2} k_x \right) \right].
\end{aligned}
\end{equation}
An external magnetic field $H_0$ aligned with the anisotropy axis can be included by adding $\tilde{\gamma} H_0$ to $\hbar \omega_a(\vec{k})$ and subtracting $\tilde{\gamma} H_0$ from $\hbar \omega_b(\vec{k})$.

The Hamiltonian Eq.~\eqref{eq:Ham} is diagonalized by a Bogoliubov transformation to the magnon modes $\hat{\alpha}$ and $\hat{\beta}$:
\begin{equation}
\frac{\hat{\mathcal{H} }_{\rm{d}}}{\hbar} = \sum_{\vec{k}}\left[ \omega_{\alpha}(\vec{k}) \hat{\alpha}^\dagger_{\vec{k}} \hat{\alpha}_{\vec{k}} +\omega_{\beta}(\vec{k}) \hat{\beta}^\dagger_{\vec{k}} \hat{\beta}_{\vec{k}} \right],
\end{equation}
where the magnon frequencies are given by
\begin{equation}
\label{eq:magnonfrequencies}
\begin{aligned}
\omega_\alpha (\vec{k}) &= \frac{\omega_a (\vec{k}) - \omega_b(\vec{k}) + \sqrt{(\omega_a (\vec{k}) + \omega_b(\vec{k}))^2 - 4 g_{ab}^2(\vec{k})}}{2}, \\
\omega_\beta (\vec{k}) &= \frac{-(\omega_a (\vec{k}) - \omega_b(\vec{k})) + \sqrt{(\omega_a (\vec{k}) + \omega_b(\vec{k}))^2 - 4 g_{ab}^2(\vec{k})}}{2}.
\end{aligned}
\end{equation}
The magnon bandgap $\Delta_{\rm{mag}} = \omega_{\alpha}(0) =\omega_{\beta}(0)$ is given explicitly by
\begin{equation}
\Delta_{\rm{mag}}=\frac{4s}{\hbar}\sqrt{16 J_1 J_{\rm{EA}} + J_{\rm{EA}}^2},
\end{equation}
while the magnon frequency splitting is
\begin{equation}
\omega_{\alpha}(\vec{k})-\omega_{\beta}(\vec{k}) = 16 s \Delta \sin \left(\frac{a}{2} k_z +\frac{a}{2} k_x \right) \sin \left(\frac{a}{2} k_z -\frac{a}{2} k_x \right).
\end{equation}

\subsection{Diffusion tensor}

In this paper, we work in the high-temperature regime, from which
$n^{0}_{\xi, \vec{k}} \approx \frac{k_B T}{\varepsilon_{\xi, \vec{k}}}$, and thus Eq.~\eqref{Eq:DiffTensor} simplifies to
\begin{equation}
\label{eq:simplifiedD}
D_{\xi, ij} =   \frac{1}{V (2 \pi)^2} \int d^2 k  \tau_{\xi, \vec{k}} v_{i,\xi \vec{k}} v_{j, \xi \vec{k}}, 
\end{equation}
where $V= \int d^2 k/(2 \pi)^2$ is the volume of the Brillouin zone. We furthermore assume that $\tau_{\xi, \vec{k}} = \tau$, i.e., the magnon momentum relaxation time is independent of the momentum and of the magnon species. Such an assumption is justified in the high-temperature regime, in which most of the magnon relaxation is to lattice phonons. At low temperatures, other processes that are neglected in our description, such as relaxation due to magnon-magnon scattering, can become relevant \cite{eto2025spontaneousmagnon}.

With the magnon bands we can evaluate the magnon velocities, and then calculate the magnon diffusion tensor in Eq.~\eqref{Eq:DiffTensor}. Even though the magnon velocities have convoluted forms, we can infer dome underlying symmetry properties that are a consequence of the symmetries of the Heisenberg Hamiltonian of Eq.~\eqref{eq:HeisHam}. Those a better analyzed in a coordinate system rotated by $\pi/4$, such that
\begin{equation}
\begin{aligned}
\frac{\sqrt{2}k_z}{2} + \frac{\sqrt{2}k_x}{2} &\rightarrow k_x , \\
\frac{\sqrt{2}k_z}{2} - \frac{\sqrt{2}k_x}{2} &\rightarrow k_z.
\end{aligned}
\end{equation}
From Eqs.~\eqref{freqs} and \eqref{eq:magnonfrequencies} we can then infer that the magnon bands are symmetric $\omega_\xi(\vec{k}) =\omega_\xi(-\vec{k})$, and related by a $\pi/2$ rotation in $k$-space:
\begin{equation}
\omega_\alpha(\vec{k}) = \omega_{\beta}(\mathcal{R}_{\pi/2} \vec{k}),
\end{equation}
corresponding to the transformation $k_x \rightarrow k_z$, and $k_z \rightarrow -k_z$. We then identify the nodal lines
\begin{equation}
\begin{aligned}
\omega_\alpha(k,0) &=\omega_\beta(k,0), \\
\omega_\alpha(0,k) &=\omega_\beta(0,k). 
\end{aligned}
\end{equation}
We can then quickly infer that $v_{x, \alpha} = v_{z, \beta} $ and $v_{z, \alpha} =  -v_{x, \beta}$, and thus $D_{\alpha, xy}= -D_{\beta, xy}= D_2$.

From the explicit form of the magnon frequencies, we further obtain that
\begin{equation}
\begin{aligned}
v_{x, \alpha} &= \mathbb{A} \cos(a^\prime k_x) \sin(a^\prime k_z) + \sin(a^\prime k_x) \mathbb{F}^{(1)}(k_x, k_z) - \sin(a^\prime k_x) \cos(a^\prime k_z) \mathbb{F}^{(2)} (k_x,k_z), \\
v_{x, \beta} &= -\mathbb{A} \cos(a^\prime k_x) \sin(a^\prime k_z) + \sin(a^\prime k_x) \mathbb{F}^{(1)}(k_x, k_z) - \sin(a^\prime k_x) \cos(a^\prime k_z) \mathbb{F}^{(2)} (k_x,k_z),\\
v_{z, \alpha} &= \mathbb{A} \cos(a^\prime k_z) \sin(a^\prime k_x) + \sin(a^\prime k_z) \mathbb{F}^{(1)}(k_x, k_z) - \sin(a^\prime k_z) \cos(a^\prime k_x) \mathbb{F}^{(2)} (k_x,k_z), \\
v_{z, \beta} &= -\mathbb{A} \cos(a^\prime k_z) \sin(a^\prime k_x) + \sin(a^\prime k_z) \mathbb{F}^{(1)}(k_x, k_z) - \sin(a^\prime k_z) \cos(a^\prime k_x) \mathbb{F}^{(2)} (k_x,k_z).
\end{aligned}
\end{equation}
where $a^\prime = a/\sqrt{2}$, and the functions $\mathbb{F}^{(1,2)}$ are independent of $\Delta$ and exhibit the following properties
\begin{equation}
\begin{aligned}
\mathbb{F}^{(1,2)}(k_x,k_z) &= \mathbb{F}^{(1,2)}(k_z,k_x), \\
\mathbb{F}^{(1,2)}(-k_x,k_z) &= \mathbb{F}^{(1,2)}(k_x,k_z) =\mathbb{F}^{(1,2)}(k_x,-k_z)=\mathbb{F}^{(1,2)}(-k_x,-k_z).
\end{aligned}
\end{equation}
For $\Delta = 0$, i.e. for an antiferromagnet, considering an integration domain that is symmetric in both $k_x$ and $k_z$, one can show by decomposing the integration in Eq.~\eqref{eq:simplifiedD} in each quadrant of the $(k_x,k_z)$ domain, that $D_2 = 0$. Thus a non-vanishing value of $D_2$, within the assumptions we have adopted, is a characteristic exclusive to ALMs. The same conclusions should be valid for general d-wave ALMs.

\section{Altermagnetic spin diffusion response function}

To obtain the spin diffusion response function, we use the magnon diffusion equations Eqs.~\eqref{eqs:DiffMag} and, where the symmetries of the ALM magnon bands, as discussed in the previous section, imply that the magnon diffusion tensors have the following form
\begin{equation}
\begin{aligned}
\overleftrightarrow{D}_\alpha = \begin{bmatrix}
    D_1 & D_2 \\
    D_2 & D_1
\end{bmatrix}, \quad \quad 
\overleftrightarrow{D}_\beta = \begin{bmatrix}
    D_1 & -D_2 \\
    -D_2 & D_1
\end{bmatrix}.
\end{aligned}
\end{equation}
We then write a diffusion equation for $\mathcal{S}_{\parallel} = \hbar (-\delta n_{\alpha} + \delta n_{\beta})$:
\begin{equation}
\label{eq:Jpara}
\partial_t \mathcal{S}_\parallel - \nabla \cdot \left[ \overleftrightarrow{D}_1 \cdot \nabla(\mathcal{S}_{\parallel} - \hbar \chi_0 h) \right] +\nabla \cdot \left[ \overleftrightarrow{D}_2 \cdot \nabla \mathcal{S}_{\perp} \right] = - \frac{1}{\tau}(\mathcal{s}_{\parallel} - \hbar \chi_0 h),
\end{equation}
where $\mathcal{S}_{\perp} = \hbar(\delta n_{\alpha} + \delta n_{\beta})$ satisfies the equation
\begin{equation}
\label{eq:Jsig}
\partial_t \mathcal{S}_\perp- \nabla \cdot \left[ \overleftrightarrow{D}_1 \cdot \nabla \mathcal{S}_{\perp} \right] +\nabla \cdot \left[ \overleftrightarrow{D}_2 \cdot \nabla (\mathcal{S}_{\parallel} - \chi_0 h) \right] = -\frac{1}{\tau} \mathcal{S}_{\perp},
\end{equation}
and we have defined
\begin{equation}
\begin{aligned}
\overleftrightarrow{D}_1 = \begin{bmatrix}
    D_1 & 0 \\
    0 & D_1
\end{bmatrix} \quad \quad 
\overleftrightarrow{D}_2 = \begin{bmatrix}
    -D_2 & 0 \\
    0 & D_2
\end{bmatrix}.
\end{aligned}
\end{equation}
Equations \eqref{eq:Jpara} and \eqref{eq:Jsig} are obtained by writting the diffusion equations in a basis that diagonalizes the magnon diffusion tensors $\overleftrightarrow{D}_{\alpha,\beta}$. We can solve this set of equations in the Fourier domain, which gives
\begin{equation}
\mathcal{S}_\parallel(\omega, \vec{k}) = \chi_\parallel(\omega, \vec{k}) h
\end{equation}
where the response function is given by Eq.~(5) of the main text:
\begin{equation}
\label{Eq:respFunctionSM}
\chi_{\parallel}(\omega, \vec{k}) = \frac{\hbar \chi_0 \left[ D_1 k^2 + \frac{1}{\tau_s} - \frac{D_2^2 k^4 \cos^2(2 \phi_k)}{- i \omega +D_1 k^2 + 1/\tau_s} \right]}{- i \omega + D_1 k^2 + \frac{1}{\tau_s} - \frac{D_2^2 k^4 \cos^2(2 \phi_k)}{- i \omega +D_1 k^2 + 1/\tau_s}},
\end{equation}
where $\phi_k = \rm{arctan}(k_x/k_z)$.

\section{Relaxation rate}

With the response function Eq.~ \eqref{Eq:respFunctionSM}, we can now calculate the QI relaxation rate with Eq.~(2) of the main text:
\begin{equation}
\label{eq:relaxSM}
\Gamma[\omega] =  \frac{\hbar \gamma^2 \tilde{\gamma}^ 2}{2}{\rm{coth}} \left( \frac{ \hbar \beta  \omega}{2} \right) \int \frac{d^2 \vec{k}}{(2 \pi)^2}  \mathcal{C}_{\{ \Phi \}}(d,\vec{k}) \chi^{\prime \prime}_{\parallel}(\omega, \vec{k}).
\end{equation}
We first recall that the geometric factor is given by $\mathcal{C}_{\{\Phi \}} (d,\vec{k}) = (2 \pi)^2 k^2 e^{- 2 k d} \mathcal{F}_{\{ \Phi \}} (\phi_k)$, where the function $\mathcal{F}_{\{ \Phi \}} (\phi_k)$ depends only on the orientation angles (see Eq.~\eqref{eq:OrienFact}), and we work in the high-temperature regime, such that:
\begin{equation}
\label{eq:relaxSM2}
\Gamma[\omega] =  \frac{ \gamma^2 \tilde{\gamma}^ 2}{\beta  \omega} \int d  \phi_k dk k^ 3 e^{- 2 k d} \mathcal{F}_{\{ \Phi \}} (\phi_k)  \chi^{\prime \prime}_{\parallel}(\omega, \vec{k}).
\end{equation}
The imaginary part of the susceptibility is given explicitly by
\begin{equation}
\chi_\parallel^ {\prime \prime} = \frac{\hbar \chi_0 \omega (D_1 k^2 + \frac{1}{\tau_s}) \left[\omega^2 + (D_1 k^2 + \frac{1}{\tau_s})^2 \right] \left[ \omega^2 + (D_1 k^2 + \frac{1}{\tau_s})^2 - D_2^2 k^4 \cos^2(2 \phi_k) \right]}{\omega^2 \left[ \omega^2 + (D_1 k^2 + \frac{1}{\tau_s})^2 + D_2^2 k^4 \cos^2(2 \phi_k) \right]^2 +  (D_1 k^2 + \frac{1}{\tau_s})^2 \left[ \omega^2 + (D_1 k^2 + \frac{1}{\tau_s})^2 - D_2^2 k^4 \cos^2(2 \phi_k) \right]^2}.
\end{equation}
We write this function in terms of the dimensionless quantities
\begin{equation}
\tilde{k} = k l_0, \quad \quad \tilde{\omega} = \omega \tau_s \quad \quad \tilde{D}_2 = D_2/D_1,
\end{equation}
where $l_0 = \sqrt{D_1 \tau_s}$, such that
\begin{equation}
\frac{\chi_\parallel^ {\prime \prime}}{\hbar \chi_0} = \tilde{\chi}_\parallel^ {\prime \prime} =  \frac{\tilde{\omega} (\tilde{k}^2 +1) \left[\tilde{\omega}^ 2 +  (\tilde{k}^2 +1)^2  \right]\left[\tilde{\omega}^ 2 +  (\tilde{k}^2 +1)^2 -\tilde{D}_2^2 \tilde{k}^4 \cos^2(2 \phi_k) \right]}{\tilde{\omega}^ 2 \left[\tilde{\omega}^ 2 +  (\tilde{k}^2 +1)^2 +\tilde{D}_2^2 \tilde{k}^4 \cos^2(2 \phi_k) \right]^ 2 + (\tilde{k}^ 2 +1)^ 2 \left[\tilde{\omega}^ 2 +  (\tilde{k}^2 +1)^2 -\tilde{D}_2^2 \tilde{k}^4 \cos^2(2 \phi_k) \right]^ 2}  .
\end{equation}
We can then write the relaxation rate as
\begin{equation}
\label{eq:relaxSM3}
\begin{aligned}
\Gamma[\omega] =  \frac{ \hbar \tau_s \chi_0 \gamma^2 \tilde{\gamma}^ 2}{\beta l_0^4  } \left[ \frac{1}{\tilde{\omega}} \int d  \phi_k d\tilde{k} \tilde{k}^ 3 e^{- 2 \tilde{k} \tilde{d}} \mathcal{F}_{\{ \Phi \}} (\phi_k)  \tilde{\chi}^{\prime \prime}_{\parallel}(\tilde{\omega}, \vec{\tilde{k}}) \right] = \Gamma_c \tilde{\Gamma}[\omega].
\end{aligned}
\end{equation}
In the above equation $\Gamma_c= \hbar k_B T \tau_s \chi_0 \gamma^2 \tilde{\gamma}^ 2 / l_0^4$ is a characteristic relaxation rate and $\tilde{d} = d/l_0$. All the plots shown in the main text are obtained via numerical integration of the above equation using standard python libraries.

\section{Spin conductivity}

To estimate the characteristic relaxation rate $\Gamma_c$, we need the static spin susceptibility $\chi_0$, which is given by the ratio $D_1/\sigma$ between the diffusion coefficient $D_1$ and the spin conductivity $\sigma$. While a full evaluation of $\sigma$ is out of scope of this paper, we can estimate it by using the procedure in \cite{wang_2022_noninvasive_measurements} for antiferromagnets. From Einstein's relation $\sigma = D_1 \partial \rho/\partial \mu$, where $\mu$ is the chemical potential and $\rho$ is the non-equilibrium magnon spin density, which can be written as
\begin{equation}
\rho = \int \frac{d^2 k}{(2 \pi)^2} \left[ \frac{1}{e^{\beta\hbar \omega_{\alpha, k} - \mu}-1} -\frac{1}{e^{\beta \hbar \omega_{\beta, k} + \mu}-1} \right].
\end{equation}
For the purposes of a simple estimate, we consider a low-momentum expansion of the magnon frequencies. We use for convenience the coordinate system of Sec. IIB
\begin{equation}
\label{eq:freqlow}
\begin{aligned}
\omega_{\alpha} &= -\Delta_{\rm{ALM}} k_x k_z + \sqrt{v^2 k^2 + \Delta^ 2}, \\
\omega_{\beta} &= \Delta_{\rm{ALM}} k_x k_z + \sqrt{v^2 k^2 + \Delta^ 2}.
\end{aligned}
\end{equation}
We do not consider the first contribution in the above expressions due to the ALM band splitting, such that
\begin{equation}
\label{eq:dens}
\frac{\partial \rho}{\partial \mu} = \frac{(k_B T)^2}{\pi^2 v^3} \left[{\rm{Li}}_2(e^{-\beta \Delta_\alpha})+ {\rm{Li}}_2(e^{-\beta \Delta_\beta}) \right],
\end{equation}
where $\Delta_{\alpha,\beta} = \Delta \pm \mu/\beta$. Typically, the chemical potential is induced by an external field which is much weaker than the magnon gap $\Delta$, as we have estimated in the main text. We thus set $\Delta_{\alpha,\beta}=\Delta$. With Eqs.~\eqref{eq:dens},\eqref{eq:freqlow} we can estimate the spin conductivity which, in units of electric conductivity, is $\sim 3.8\times 10^6$ S/m for the parameters used in the main text and at $200$ K.

The calculations presented here are a rough estimate of the spin conductivity used as a starting point to assess the feasibility of our scheme. A more complete description can be done within the framework of, e.g. \cite{fang_2022_generalizedmodel}, which also accommodates the ballistic spin transport regime. An extension to such a complete framework is postponed to a future work.

\bibliographystyle{apsrev4-1}
\bibliography{biblioaltermagnon} 

\begin{thebibliography}{78}%
\makeatletter
\providecommand \@ifxundefined [1]{%
 \@ifx{#1\undefined}
}%
\providecommand \@ifnum [1]{%
 \ifnum #1\expandafter \@firstoftwo
 \else \expandafter \@secondoftwo
 \fi
}%
\providecommand \@ifx [1]{%
 \ifx #1\expandafter \@firstoftwo
 \else \expandafter \@secondoftwo
 \fi
}%
\providecommand \natexlab [1]{#1}%
\providecommand \enquote  [1]{``#1''}%
\providecommand \bibnamefont  [1]{#1}%
\providecommand \bibfnamefont [1]{#1}%
\providecommand \citenamefont [1]{#1}%
\providecommand \href@noop [0]{\@secondoftwo}%
\providecommand \href [0]{\begingroup \@sanitize@url \@href}%
\providecommand \@href[1]{\@@startlink{#1}\@@href}%
\providecommand \@@href[1]{\endgroup#1\@@endlink}%
\providecommand \@sanitize@url [0]{\catcode `\\12\catcode `\$12\catcode
  `\&12\catcode `\#12\catcode `\^12\catcode `\_12\catcode `\%12\relax}%
\providecommand \@@startlink[1]{}%
\providecommand \@@endlink[0]{}%
\providecommand \url  [0]{\begingroup\@sanitize@url \@url }%
\providecommand \@url [1]{\endgroup\@href {#1}{\urlprefix }}%
\providecommand \urlprefix  [0]{URL }%
\providecommand \Eprint [0]{\href }%
\providecommand \doibase [0]{http://dx.doi.org/}%
\providecommand \selectlanguage [0]{\@gobble}%
\providecommand \bibinfo  [0]{\@secondoftwo}%
\providecommand \bibfield  [0]{\@secondoftwo}%
\providecommand \translation [1]{[#1]}%
\providecommand \BibitemOpen [0]{}%
\providecommand \bibitemStop [0]{}%
\providecommand \bibitemNoStop [0]{.\EOS\space}%
\providecommand \EOS [0]{\spacefactor3000\relax}%
\providecommand \BibitemShut  [1]{\csname bibitem#1\endcsname}%
\let\auto@bib@innerbib\@empty
\bibitem [{\citenamefont {\ifmmode~\check{S}\else \v{S}\fi{}mejkal}\ \emph
  {et~al.}(2022{\natexlab{a}})\citenamefont {\ifmmode~\check{S}\else
  \v{S}\fi{}mejkal}, \citenamefont {Sinova},\ and\ \citenamefont
  {Jungwirth}}]{PhysRevX.12.031042}%
  \BibitemOpen
  \bibfield  {author} {\bibinfo {author} {\bibfnamefont {L.}~\bibnamefont
  {\ifmmode~\check{S}\else \v{S}\fi{}mejkal}}, \bibinfo {author} {\bibfnamefont
  {J.}~\bibnamefont {Sinova}}, \ and\ \bibinfo {author} {\bibfnamefont
  {T.}~\bibnamefont {Jungwirth}},\ }\href {\doibase 10.1103/PhysRevX.12.031042}
  {\bibfield  {journal} {\bibinfo  {journal} {Phys. Rev. X}\ }\textbf {\bibinfo
  {volume} {12}},\ \bibinfo {pages} {031042} (\bibinfo {year}
  {2022}{\natexlab{a}})}\BibitemShut {NoStop}%
\bibitem [{\citenamefont {\ifmmode~\check{S}\else \v{S}\fi{}mejkal}\ \emph
  {et~al.}(2022{\natexlab{b}})\citenamefont {\ifmmode~\check{S}\else
  \v{S}\fi{}mejkal}, \citenamefont {Sinova},\ and\ \citenamefont
  {Jungwirth}}]{PhysRevX.12.040501}%
  \BibitemOpen
  \bibfield  {author} {\bibinfo {author} {\bibfnamefont {L.}~\bibnamefont
  {\ifmmode~\check{S}\else \v{S}\fi{}mejkal}}, \bibinfo {author} {\bibfnamefont
  {J.}~\bibnamefont {Sinova}}, \ and\ \bibinfo {author} {\bibfnamefont
  {T.}~\bibnamefont {Jungwirth}},\ }\href {\doibase 10.1103/PhysRevX.12.040501}
  {\bibfield  {journal} {\bibinfo  {journal} {Phys. Rev. X}\ }\textbf {\bibinfo
  {volume} {12}},\ \bibinfo {pages} {040501} (\bibinfo {year}
  {2022}{\natexlab{b}})}\BibitemShut {NoStop}%
\bibitem [{\citenamefont {Jungwirth}\ \emph {et~al.}(2025)\citenamefont
  {Jungwirth}, \citenamefont {Fernandes}, \citenamefont {Fradkin},
  \citenamefont {MacDonald}, \citenamefont {Sinova},\ and\ \citenamefont
  {{\v{S}}mejkal}}]{Jungwirth_2025_altermagnetism}%
  \BibitemOpen
  \bibfield  {author} {\bibinfo {author} {\bibfnamefont {T.}~\bibnamefont
  {Jungwirth}}, \bibinfo {author} {\bibfnamefont {R.~M.}\ \bibnamefont
  {Fernandes}}, \bibinfo {author} {\bibfnamefont {E.}~\bibnamefont {Fradkin}},
  \bibinfo {author} {\bibfnamefont {A.~H.}\ \bibnamefont {MacDonald}}, \bibinfo
  {author} {\bibfnamefont {J.}~\bibnamefont {Sinova}}, \ and\ \bibinfo {author}
  {\bibfnamefont {L.}~\bibnamefont {{\v{S}}mejkal}},\ }\href
  {https://doi.org/10.1016/j.newton.2025.100162} {\bibfield  {journal}
  {\bibinfo  {journal} {Newton}\ } (\bibinfo {year} {2025})}\BibitemShut
  {NoStop}%
\bibitem [{\citenamefont
  {Mazin}(2022{\natexlab{a}})}]{mazin_2022_aletmagnetism}%
  \BibitemOpen
  \bibfield  {author} {\bibinfo {author} {\bibfnamefont {I.}~\bibnamefont
  {Mazin}} (\bibinfo {collaboration} {The PRX Editors}),\ }\href {\doibase
  10.1103/PhysRevX.12.040002} {\bibfield  {journal} {\bibinfo  {journal} {Phys.
  Rev. X}\ }\textbf {\bibinfo {volume} {12}},\ \bibinfo {pages} {040002}
  (\bibinfo {year} {2022}{\natexlab{a}})}\BibitemShut {NoStop}%
\bibitem [{\citenamefont {Bai}\ \emph {et~al.}(2024)\citenamefont {Bai},
  \citenamefont {Feng}, \citenamefont {Liu}, \citenamefont {Šmejkal},
  \citenamefont {Mokrousov},\ and\ \citenamefont
  {Yao}}]{bai_2024_altermagnetism}%
  \BibitemOpen
  \bibfield  {author} {\bibinfo {author} {\bibfnamefont {L.}~\bibnamefont
  {Bai}}, \bibinfo {author} {\bibfnamefont {W.}~\bibnamefont {Feng}}, \bibinfo
  {author} {\bibfnamefont {S.}~\bibnamefont {Liu}}, \bibinfo {author}
  {\bibfnamefont {L.}~\bibnamefont {Šmejkal}}, \bibinfo {author}
  {\bibfnamefont {Y.}~\bibnamefont {Mokrousov}}, \ and\ \bibinfo {author}
  {\bibfnamefont {Y.}~\bibnamefont {Yao}},\ }\href {\doibase
  https://doi.org/10.1002/adfm.202409327} {\bibfield  {journal} {\bibinfo
  {journal} {Advanced Functional Materials}\ }\textbf {\bibinfo {volume}
  {34}},\ \bibinfo {pages} {2409327} (\bibinfo {year} {2024})}\BibitemShut
  {NoStop}%
\bibitem [{\citenamefont {Mazin}\ \emph {et~al.}(2023)\citenamefont {Mazin},
  \citenamefont {Gonz{\'a}lez-Hern{\'a}ndez},\ and\ \citenamefont
  {{\v{S}}mejkal}}]{mazin2023induced}%
  \BibitemOpen
  \bibfield  {author} {\bibinfo {author} {\bibfnamefont {I.}~\bibnamefont
  {Mazin}}, \bibinfo {author} {\bibfnamefont {R.}~\bibnamefont
  {Gonz{\'a}lez-Hern{\'a}ndez}}, \ and\ \bibinfo {author} {\bibfnamefont
  {L.}~\bibnamefont {{\v{S}}mejkal}},\ }\href
  {https://arxiv.org/abs/2309.02355} {\bibfield  {journal} {\bibinfo  {journal}
  {arXiv preprint arXiv:2309.02355}\ } (\bibinfo {year} {2023})}\BibitemShut
  {NoStop}%
\bibitem [{\citenamefont {{\v{S}}mejkal}\ \emph {et~al.}(2024)\citenamefont
  {{\v{S}}mejkal}, \citenamefont {D’souza}, \citenamefont {Hajlaoui},
  \citenamefont {Springholz}, \citenamefont {Uhl{\'\i}{\v{r}}ov{\'a}},
  \citenamefont {Alarab}, \citenamefont {Constantinou}, \citenamefont
  {Strocov}, \citenamefont {Usanov} \emph
  {et~al.}}]{krempasky2024altermagnetic}%
  \BibitemOpen
  \bibfield  {author} {\bibinfo {author} {\bibfnamefont {L.}~\bibnamefont
  {{\v{S}}mejkal}}, \bibinfo {author} {\bibfnamefont {S.}~\bibnamefont
  {D’souza}}, \bibinfo {author} {\bibfnamefont {M.}~\bibnamefont {Hajlaoui}},
  \bibinfo {author} {\bibfnamefont {G.}~\bibnamefont {Springholz}}, \bibinfo
  {author} {\bibfnamefont {K.}~\bibnamefont {Uhl{\'\i}{\v{r}}ov{\'a}}},
  \bibinfo {author} {\bibfnamefont {F.}~\bibnamefont {Alarab}}, \bibinfo
  {author} {\bibfnamefont {P.}~\bibnamefont {Constantinou}}, \bibinfo {author}
  {\bibfnamefont {V.}~\bibnamefont {Strocov}}, \bibinfo {author} {\bibfnamefont
  {D.}~\bibnamefont {Usanov}},  \emph {et~al.},\ }\href
  {https://www.nature.com/articles/s41586-023-06907-7} {\bibfield  {journal}
  {\bibinfo  {journal} {Nature}\ }\textbf {\bibinfo {volume} {626}},\ \bibinfo
  {pages} {517} (\bibinfo {year} {2024})}\BibitemShut {NoStop}%
\bibitem [{\citenamefont {Fedchenko}\ \emph {et~al.}(2024)\citenamefont
  {Fedchenko}, \citenamefont {Min{\'a}r}, \citenamefont {Akashdeep},
  \citenamefont {D’Souza}, \citenamefont {Vasilyev}, \citenamefont {Tkach},
  \citenamefont {Odenbreit}, \citenamefont {Nguyen}, \citenamefont
  {Kutnyakhov}, \citenamefont {Wind} \emph
  {et~al.}}]{fedchenko2024observation}%
  \BibitemOpen
  \bibfield  {author} {\bibinfo {author} {\bibfnamefont {O.}~\bibnamefont
  {Fedchenko}}, \bibinfo {author} {\bibfnamefont {J.}~\bibnamefont
  {Min{\'a}r}}, \bibinfo {author} {\bibfnamefont {A.}~\bibnamefont
  {Akashdeep}}, \bibinfo {author} {\bibfnamefont {S.~W.}\ \bibnamefont
  {D’Souza}}, \bibinfo {author} {\bibfnamefont {D.}~\bibnamefont {Vasilyev}},
  \bibinfo {author} {\bibfnamefont {O.}~\bibnamefont {Tkach}}, \bibinfo
  {author} {\bibfnamefont {L.}~\bibnamefont {Odenbreit}}, \bibinfo {author}
  {\bibfnamefont {Q.}~\bibnamefont {Nguyen}}, \bibinfo {author} {\bibfnamefont
  {D.}~\bibnamefont {Kutnyakhov}}, \bibinfo {author} {\bibfnamefont
  {N.}~\bibnamefont {Wind}},  \emph {et~al.},\ }\href
  {https://www.science.org/doi/10.1126/sciadv.adj4883} {\bibfield  {journal}
  {\bibinfo  {journal} {Science advances}\ }\textbf {\bibinfo {volume} {10}},\
  \bibinfo {pages} {eadj4883} (\bibinfo {year} {2024})}\BibitemShut {NoStop}%
\bibitem [{\citenamefont {Reimers}\ \emph {et~al.}(2024)\citenamefont
  {Reimers}, \citenamefont {Odenbreit}, \citenamefont {{\v{S}}mejkal},
  \citenamefont {Strocov}, \citenamefont {Constantinou}, \citenamefont
  {Hellenes}, \citenamefont {Jaeschke~Ubiergo}, \citenamefont {Campos},
  \citenamefont {Bharadwaj}, \citenamefont {Chakraborty} \emph
  {et~al.}}]{reimers2024direct}%
  \BibitemOpen
  \bibfield  {author} {\bibinfo {author} {\bibfnamefont {S.}~\bibnamefont
  {Reimers}}, \bibinfo {author} {\bibfnamefont {L.}~\bibnamefont {Odenbreit}},
  \bibinfo {author} {\bibfnamefont {L.}~\bibnamefont {{\v{S}}mejkal}}, \bibinfo
  {author} {\bibfnamefont {V.~N.}\ \bibnamefont {Strocov}}, \bibinfo {author}
  {\bibfnamefont {P.}~\bibnamefont {Constantinou}}, \bibinfo {author}
  {\bibfnamefont {A.~B.}\ \bibnamefont {Hellenes}}, \bibinfo {author}
  {\bibfnamefont {R.}~\bibnamefont {Jaeschke~Ubiergo}}, \bibinfo {author}
  {\bibfnamefont {W.~H.}\ \bibnamefont {Campos}}, \bibinfo {author}
  {\bibfnamefont {V.~K.}\ \bibnamefont {Bharadwaj}}, \bibinfo {author}
  {\bibfnamefont {A.}~\bibnamefont {Chakraborty}},  \emph {et~al.},\ }\href
  {https://www.nature.com/articles/s41467-024-46476-5} {\bibfield  {journal}
  {\bibinfo  {journal} {Nature Communications}\ }\textbf {\bibinfo {volume}
  {15}},\ \bibinfo {pages} {2116} (\bibinfo {year} {2024})}\BibitemShut
  {NoStop}%
\bibitem [{\citenamefont {Lee}\ \emph {et~al.}(2024)\citenamefont {Lee},
  \citenamefont {Lee}, \citenamefont {Jung}, \citenamefont {Jung},
  \citenamefont {Kim}, \citenamefont {Lee}, \citenamefont {Seok}, \citenamefont
  {Kim}, \citenamefont {Park}, \citenamefont {\ifmmode~\check{S}\else
  \v{S}\fi{}mejkal}, \citenamefont {Kang},\ and\ \citenamefont
  {Kim}}]{lee_2024_broken}%
  \BibitemOpen
  \bibfield  {author} {\bibinfo {author} {\bibfnamefont {S.}~\bibnamefont
  {Lee}}, \bibinfo {author} {\bibfnamefont {S.}~\bibnamefont {Lee}}, \bibinfo
  {author} {\bibfnamefont {S.}~\bibnamefont {Jung}}, \bibinfo {author}
  {\bibfnamefont {J.}~\bibnamefont {Jung}}, \bibinfo {author} {\bibfnamefont
  {D.}~\bibnamefont {Kim}}, \bibinfo {author} {\bibfnamefont {Y.}~\bibnamefont
  {Lee}}, \bibinfo {author} {\bibfnamefont {B.}~\bibnamefont {Seok}}, \bibinfo
  {author} {\bibfnamefont {J.}~\bibnamefont {Kim}}, \bibinfo {author}
  {\bibfnamefont {B.~G.}\ \bibnamefont {Park}}, \bibinfo {author}
  {\bibfnamefont {L.}~\bibnamefont {\ifmmode~\check{S}\else \v{S}\fi{}mejkal}},
  \bibinfo {author} {\bibfnamefont {C.-J.}\ \bibnamefont {Kang}}, \ and\
  \bibinfo {author} {\bibfnamefont {C.}~\bibnamefont {Kim}},\ }\href {\doibase
  10.1103/PhysRevLett.132.036702} {\bibfield  {journal} {\bibinfo  {journal}
  {Phys. Rev. Lett.}\ }\textbf {\bibinfo {volume} {132}},\ \bibinfo {pages}
  {036702} (\bibinfo {year} {2024})}\BibitemShut {NoStop}%
\bibitem [{\citenamefont {Jiang}\ \emph {et~al.}(2025)\citenamefont {Jiang},
  \citenamefont {Hu}, \citenamefont {Bai}, \citenamefont {Song}, \citenamefont
  {Mu}, \citenamefont {Qu}, \citenamefont {Li}, \citenamefont {Zhu},
  \citenamefont {Pi}, \citenamefont {Wei}, \citenamefont {Sun}, \citenamefont
  {Huang}, \citenamefont {Zheng}, \citenamefont {Peng}, \citenamefont {He},
  \citenamefont {Li}, \citenamefont {Luo}, \citenamefont {Li}, \citenamefont
  {Chen}, \citenamefont {Li}, \citenamefont {Weng},\ and\ \citenamefont
  {Qian}}]{Jiang2025}%
  \BibitemOpen
  \bibfield  {author} {\bibinfo {author} {\bibfnamefont {B.}~\bibnamefont
  {Jiang}}, \bibinfo {author} {\bibfnamefont {M.}~\bibnamefont {Hu}}, \bibinfo
  {author} {\bibfnamefont {J.}~\bibnamefont {Bai}}, \bibinfo {author}
  {\bibfnamefont {Z.}~\bibnamefont {Song}}, \bibinfo {author} {\bibfnamefont
  {C.}~\bibnamefont {Mu}}, \bibinfo {author} {\bibfnamefont {G.}~\bibnamefont
  {Qu}}, \bibinfo {author} {\bibfnamefont {W.}~\bibnamefont {Li}}, \bibinfo
  {author} {\bibfnamefont {W.}~\bibnamefont {Zhu}}, \bibinfo {author}
  {\bibfnamefont {H.}~\bibnamefont {Pi}}, \bibinfo {author} {\bibfnamefont
  {Z.}~\bibnamefont {Wei}}, \bibinfo {author} {\bibfnamefont {Y.-J.}\
  \bibnamefont {Sun}}, \bibinfo {author} {\bibfnamefont {Y.}~\bibnamefont
  {Huang}}, \bibinfo {author} {\bibfnamefont {X.}~\bibnamefont {Zheng}},
  \bibinfo {author} {\bibfnamefont {Y.}~\bibnamefont {Peng}}, \bibinfo {author}
  {\bibfnamefont {L.}~\bibnamefont {He}}, \bibinfo {author} {\bibfnamefont
  {S.}~\bibnamefont {Li}}, \bibinfo {author} {\bibfnamefont {J.}~\bibnamefont
  {Luo}}, \bibinfo {author} {\bibfnamefont {Z.}~\bibnamefont {Li}}, \bibinfo
  {author} {\bibfnamefont {G.}~\bibnamefont {Chen}}, \bibinfo {author}
  {\bibfnamefont {H.}~\bibnamefont {Li}}, \bibinfo {author} {\bibfnamefont
  {H.}~\bibnamefont {Weng}}, \ and\ \bibinfo {author} {\bibfnamefont
  {T.}~\bibnamefont {Qian}},\ }\href {\doibase 10.1038/s41567-025-02822-y}
  {\bibfield  {journal} {\bibinfo  {journal} {Nature Physics}\ }\textbf
  {\bibinfo {volume} {21}},\ \bibinfo {pages} {754} (\bibinfo {year}
  {2025})}\BibitemShut {NoStop}%
\bibitem [{\citenamefont {Zhang}\ \emph {et~al.}(2025)\citenamefont {Zhang},
  \citenamefont {Cheng}, \citenamefont {Yin}, \citenamefont {Liu},
  \citenamefont {Deng}, \citenamefont {Qiao}, \citenamefont {Shi},
  \citenamefont {Zhang}, \citenamefont {Lin}, \citenamefont {Liu},
  \citenamefont {Ye}, \citenamefont {Huang}, \citenamefont {Meng},
  \citenamefont {Zhang}, \citenamefont {Okuda}, \citenamefont {Shimada},
  \citenamefont {Cui}, \citenamefont {Zhao}, \citenamefont {Cao}, \citenamefont
  {Qiao}, \citenamefont {Liu},\ and\ \citenamefont {Chen}}]{Zhang2025}%
  \BibitemOpen
  \bibfield  {author} {\bibinfo {author} {\bibfnamefont {F.}~\bibnamefont
  {Zhang}}, \bibinfo {author} {\bibfnamefont {X.}~\bibnamefont {Cheng}},
  \bibinfo {author} {\bibfnamefont {Z.}~\bibnamefont {Yin}}, \bibinfo {author}
  {\bibfnamefont {C.}~\bibnamefont {Liu}}, \bibinfo {author} {\bibfnamefont
  {L.}~\bibnamefont {Deng}}, \bibinfo {author} {\bibfnamefont {Y.}~\bibnamefont
  {Qiao}}, \bibinfo {author} {\bibfnamefont {Z.}~\bibnamefont {Shi}}, \bibinfo
  {author} {\bibfnamefont {S.}~\bibnamefont {Zhang}}, \bibinfo {author}
  {\bibfnamefont {J.}~\bibnamefont {Lin}}, \bibinfo {author} {\bibfnamefont
  {Z.}~\bibnamefont {Liu}}, \bibinfo {author} {\bibfnamefont {M.}~\bibnamefont
  {Ye}}, \bibinfo {author} {\bibfnamefont {Y.}~\bibnamefont {Huang}}, \bibinfo
  {author} {\bibfnamefont {X.}~\bibnamefont {Meng}}, \bibinfo {author}
  {\bibfnamefont {C.}~\bibnamefont {Zhang}}, \bibinfo {author} {\bibfnamefont
  {T.}~\bibnamefont {Okuda}}, \bibinfo {author} {\bibfnamefont
  {K.}~\bibnamefont {Shimada}}, \bibinfo {author} {\bibfnamefont
  {S.}~\bibnamefont {Cui}}, \bibinfo {author} {\bibfnamefont {Y.}~\bibnamefont
  {Zhao}}, \bibinfo {author} {\bibfnamefont {G.-H.}\ \bibnamefont {Cao}},
  \bibinfo {author} {\bibfnamefont {S.}~\bibnamefont {Qiao}}, \bibinfo {author}
  {\bibfnamefont {J.}~\bibnamefont {Liu}}, \ and\ \bibinfo {author}
  {\bibfnamefont {C.}~\bibnamefont {Chen}},\ }\href {\doibase
  10.1038/s41567-025-02864-2} {\bibfield  {journal} {\bibinfo  {journal}
  {Nature Physics}\ }\textbf {\bibinfo {volume} {21}},\ \bibinfo {pages} {760}
  (\bibinfo {year} {2025})}\BibitemShut {NoStop}%
\bibitem [{\citenamefont {{\v{S}}mejkal}\ \emph {et~al.}(2020)\citenamefont
  {{\v{S}}mejkal}, \citenamefont {Gonz\'{a}lez-Hern\'{a}ndez}, \citenamefont
  {Jungwirth},\ and\ \citenamefont {Sinova}}]{libor_2020}%
  \BibitemOpen
  \bibfield  {author} {\bibinfo {author} {\bibfnamefont {L.}~\bibnamefont
  {{\v{S}}mejkal}}, \bibinfo {author} {\bibfnamefont {R.}~\bibnamefont
  {Gonz\'{a}lez-Hern\'{a}ndez}}, \bibinfo {author} {\bibfnamefont
  {T.}~\bibnamefont {Jungwirth}}, \ and\ \bibinfo {author} {\bibfnamefont
  {J.}~\bibnamefont {Sinova}},\ }\href {\doibase 10.1126/sciadv.aaz8809}
  {\bibfield  {journal} {\bibinfo  {journal} {Science Advances}\ }\textbf
  {\bibinfo {volume} {6}},\ \bibinfo {pages} {eaaz8809} (\bibinfo {year}
  {2020})}\BibitemShut {NoStop}%
\bibitem [{\citenamefont {Mazin}\ \emph {et~al.}(2021)\citenamefont {Mazin},
  \citenamefont {Koepernik}, \citenamefont {Johannes}, \citenamefont
  {Gonz\'{a}lez-Hern\'{a}ndez},\ and\ \citenamefont
  {{\v{S}}mejkal}}]{mazin_2021_prediction}%
  \BibitemOpen
  \bibfield  {author} {\bibinfo {author} {\bibfnamefont {I.~I.}\ \bibnamefont
  {Mazin}}, \bibinfo {author} {\bibfnamefont {K.}~\bibnamefont {Koepernik}},
  \bibinfo {author} {\bibfnamefont {M.~D.}\ \bibnamefont {Johannes}}, \bibinfo
  {author} {\bibfnamefont {R.}~\bibnamefont {Gonz\'{a}lez-Hern\'{a}ndez}}, \
  and\ \bibinfo {author} {\bibfnamefont {L.}~\bibnamefont {{\v{S}}mejkal}},\
  }\href {\doibase 10.1073/pnas.2108924118} {\bibfield  {journal} {\bibinfo
  {journal} {Proceedings of the National Academy of Sciences}\ }\textbf
  {\bibinfo {volume} {118}},\ \bibinfo {pages} {e2108924118} (\bibinfo {year}
  {2021})}\BibitemShut {NoStop}%
\bibitem [{\citenamefont {Feng}\ \emph {et~al.}(2022)\citenamefont {Feng},
  \citenamefont {Zhou}, \citenamefont {{\v{S}}mejkal}, \citenamefont {Wu},
  \citenamefont {Zhu}, \citenamefont {Guo}, \citenamefont
  {Gonz{\'a}lez-Hern{\'a}ndez}, \citenamefont {Wang}, \citenamefont {Yan},
  \citenamefont {Qin}, \citenamefont {Zhang}, \citenamefont {Wu}, \citenamefont
  {Chen}, \citenamefont {Meng}, \citenamefont {Liu}, \citenamefont {Xia},
  \citenamefont {Sinova}, \citenamefont {Jungwirth},\ and\ \citenamefont
  {Liu}}]{Feng2022}%
  \BibitemOpen
  \bibfield  {author} {\bibinfo {author} {\bibfnamefont {Z.}~\bibnamefont
  {Feng}}, \bibinfo {author} {\bibfnamefont {X.}~\bibnamefont {Zhou}}, \bibinfo
  {author} {\bibfnamefont {L.}~\bibnamefont {{\v{S}}mejkal}}, \bibinfo {author}
  {\bibfnamefont {L.}~\bibnamefont {Wu}}, \bibinfo {author} {\bibfnamefont
  {Z.}~\bibnamefont {Zhu}}, \bibinfo {author} {\bibfnamefont {H.}~\bibnamefont
  {Guo}}, \bibinfo {author} {\bibfnamefont {R.}~\bibnamefont
  {Gonz{\'a}lez-Hern{\'a}ndez}}, \bibinfo {author} {\bibfnamefont
  {X.}~\bibnamefont {Wang}}, \bibinfo {author} {\bibfnamefont {H.}~\bibnamefont
  {Yan}}, \bibinfo {author} {\bibfnamefont {P.}~\bibnamefont {Qin}}, \bibinfo
  {author} {\bibfnamefont {X.}~\bibnamefont {Zhang}}, \bibinfo {author}
  {\bibfnamefont {H.}~\bibnamefont {Wu}}, \bibinfo {author} {\bibfnamefont
  {H.}~\bibnamefont {Chen}}, \bibinfo {author} {\bibfnamefont {Z.}~\bibnamefont
  {Meng}}, \bibinfo {author} {\bibfnamefont {L.}~\bibnamefont {Liu}}, \bibinfo
  {author} {\bibfnamefont {Z.}~\bibnamefont {Xia}}, \bibinfo {author}
  {\bibfnamefont {J.}~\bibnamefont {Sinova}}, \bibinfo {author} {\bibfnamefont
  {T.}~\bibnamefont {Jungwirth}}, \ and\ \bibinfo {author} {\bibfnamefont
  {Z.}~\bibnamefont {Liu}},\ }\href {\doibase 10.1038/s41928-022-00866-z}
  {\bibfield  {journal} {\bibinfo  {journal} {Nature Electronics}\ }\textbf
  {\bibinfo {volume} {5}},\ \bibinfo {pages} {735} (\bibinfo {year}
  {2022})}\BibitemShut {NoStop}%
\bibitem [{\citenamefont {Reichlova}\ \emph {et~al.}(2024)\citenamefont
  {Reichlova}, \citenamefont {Lopes~Seeger}, \citenamefont
  {Gonz{\'a}lez-Hern{\'a}ndez}, \citenamefont {Kounta}, \citenamefont
  {Schlitz}, \citenamefont {Kriegner}, \citenamefont {Ritzinger}, \citenamefont
  {Lammel}, \citenamefont {Leivisk{\"a}}, \citenamefont {Birk~Hellenes} \emph
  {et~al.}}]{reichlova2024observation}%
  \BibitemOpen
  \bibfield  {author} {\bibinfo {author} {\bibfnamefont {H.}~\bibnamefont
  {Reichlova}}, \bibinfo {author} {\bibfnamefont {R.}~\bibnamefont
  {Lopes~Seeger}}, \bibinfo {author} {\bibfnamefont {R.}~\bibnamefont
  {Gonz{\'a}lez-Hern{\'a}ndez}}, \bibinfo {author} {\bibfnamefont
  {I.}~\bibnamefont {Kounta}}, \bibinfo {author} {\bibfnamefont
  {R.}~\bibnamefont {Schlitz}}, \bibinfo {author} {\bibfnamefont
  {D.}~\bibnamefont {Kriegner}}, \bibinfo {author} {\bibfnamefont
  {P.}~\bibnamefont {Ritzinger}}, \bibinfo {author} {\bibfnamefont
  {M.}~\bibnamefont {Lammel}}, \bibinfo {author} {\bibfnamefont
  {M.}~\bibnamefont {Leivisk{\"a}}}, \bibinfo {author} {\bibfnamefont
  {A.}~\bibnamefont {Birk~Hellenes}},  \emph {et~al.},\ }\href
  {https://www.nature.com/articles/s41467-024-48493-w} {\bibfield  {journal}
  {\bibinfo  {journal} {Nature Communications}\ }\textbf {\bibinfo {volume}
  {15}},\ \bibinfo {pages} {4961} (\bibinfo {year} {2024})}\BibitemShut
  {NoStop}%
\bibitem [{\citenamefont {\ifmmode~\check{S}\else \v{S}\fi{}mejkal}\ \emph
  {et~al.}(2022{\natexlab{c}})\citenamefont {\ifmmode~\check{S}\else
  \v{S}\fi{}mejkal}, \citenamefont {Hellenes}, \citenamefont
  {Gonz\'alez-Hern\'andez}, \citenamefont {Sinova},\ and\ \citenamefont
  {Jungwirth}}]{smejkal_2022_giant_tunneling}%
  \BibitemOpen
  \bibfield  {author} {\bibinfo {author} {\bibfnamefont {L.}~\bibnamefont
  {\ifmmode~\check{S}\else \v{S}\fi{}mejkal}}, \bibinfo {author} {\bibfnamefont
  {A.~B.}\ \bibnamefont {Hellenes}}, \bibinfo {author} {\bibfnamefont
  {R.}~\bibnamefont {Gonz\'alez-Hern\'andez}}, \bibinfo {author} {\bibfnamefont
  {J.}~\bibnamefont {Sinova}}, \ and\ \bibinfo {author} {\bibfnamefont
  {T.}~\bibnamefont {Jungwirth}},\ }\href {\doibase 10.1103/PhysRevX.12.011028}
  {\bibfield  {journal} {\bibinfo  {journal} {Phys. Rev. X}\ }\textbf {\bibinfo
  {volume} {12}},\ \bibinfo {pages} {011028} (\bibinfo {year}
  {2022}{\natexlab{c}})}\BibitemShut {NoStop}%
\bibitem [{\citenamefont {Antonenko}\ \emph {et~al.}(2025)\citenamefont
  {Antonenko}, \citenamefont {Fernandes},\ and\ \citenamefont
  {Venderbos}}]{PhysRevLett.134.096703}%
  \BibitemOpen
  \bibfield  {author} {\bibinfo {author} {\bibfnamefont {D.~S.}\ \bibnamefont
  {Antonenko}}, \bibinfo {author} {\bibfnamefont {R.~M.}\ \bibnamefont
  {Fernandes}}, \ and\ \bibinfo {author} {\bibfnamefont {J.~W.~F.}\
  \bibnamefont {Venderbos}},\ }\href {\doibase 10.1103/PhysRevLett.134.096703}
  {\bibfield  {journal} {\bibinfo  {journal} {Phys. Rev. Lett.}\ }\textbf
  {\bibinfo {volume} {134}},\ \bibinfo {pages} {096703} (\bibinfo {year}
  {2025})}\BibitemShut {NoStop}%
\bibitem [{\citenamefont {J.~W.~González}(2025)}]{Gonzalez2025Dual}%
  \BibitemOpen
  \bibfield  {author} {\bibinfo {author} {\bibfnamefont {N.~V.-S. A. M.~L.}\
  \bibnamefont {J.~W.~González}, \bibfnamefont {R.~A.~Gallardo}},\ }\href
  {https://arxiv.org/abs/2508.15944} {\bibfield  {journal} {\bibinfo  {journal}
  {arXiv preprint arXiv:2508.15944}\ } (\bibinfo {year} {2025})}\BibitemShut
  {NoStop}%
\bibitem [{\citenamefont {Duan}\ \emph
  {et~al.}(2025{\natexlab{a}})\citenamefont {Duan}, \citenamefont {Zhang},
  \citenamefont {Zhu}, \citenamefont {Liu}, \citenamefont {Zhang},
  \citenamefont {\ifmmode \check{Z}\else \v{Z}\fi{}uti\ifmmode~\acute{c}\else
  \'{c}\fi{}},\ and\ \citenamefont {Zhou}}]{PhysRevLett.134.106801}%
  \BibitemOpen
  \bibfield  {author} {\bibinfo {author} {\bibfnamefont {X.}~\bibnamefont
  {Duan}}, \bibinfo {author} {\bibfnamefont {J.}~\bibnamefont {Zhang}},
  \bibinfo {author} {\bibfnamefont {Z.}~\bibnamefont {Zhu}}, \bibinfo {author}
  {\bibfnamefont {Y.}~\bibnamefont {Liu}}, \bibinfo {author} {\bibfnamefont
  {Z.}~\bibnamefont {Zhang}}, \bibinfo {author} {\bibfnamefont
  {I.}~\bibnamefont {\ifmmode \check{Z}\else
  \v{Z}\fi{}uti\ifmmode~\acute{c}\else \'{c}\fi{}}}, \ and\ \bibinfo {author}
  {\bibfnamefont {T.}~\bibnamefont {Zhou}},\ }\href {\doibase
  10.1103/PhysRevLett.134.106801} {\bibfield  {journal} {\bibinfo  {journal}
  {Phys. Rev. Lett.}\ }\textbf {\bibinfo {volume} {134}},\ \bibinfo {pages}
  {106801} (\bibinfo {year} {2025}{\natexlab{a}})}\BibitemShut {NoStop}%
\bibitem [{\citenamefont {{\v{S}}mejkal}(2024)}]{libormultiferroics2024}%
  \BibitemOpen
  \bibfield  {author} {\bibinfo {author} {\bibfnamefont {L.}~\bibnamefont
  {{\v{S}}mejkal}},\ }\href@noop {} {\bibfield  {journal} {\bibinfo  {journal}
  {arXiv preprint arXiv:2411.19928}\ } (\bibinfo {year} {2024})}\BibitemShut
  {NoStop}%
\bibitem [{\citenamefont {Gu}\ \emph {et~al.}(2025)\citenamefont {Gu},
  \citenamefont {Liu}, \citenamefont {Zhu}, \citenamefont {Yananose},
  \citenamefont {Chen}, \citenamefont {Hu}, \citenamefont {Stroppa},\ and\
  \citenamefont {Liu}}]{PhysRevLett.134.106802}%
  \BibitemOpen
  \bibfield  {author} {\bibinfo {author} {\bibfnamefont {M.}~\bibnamefont
  {Gu}}, \bibinfo {author} {\bibfnamefont {Y.}~\bibnamefont {Liu}}, \bibinfo
  {author} {\bibfnamefont {H.}~\bibnamefont {Zhu}}, \bibinfo {author}
  {\bibfnamefont {K.}~\bibnamefont {Yananose}}, \bibinfo {author}
  {\bibfnamefont {X.}~\bibnamefont {Chen}}, \bibinfo {author} {\bibfnamefont
  {Y.}~\bibnamefont {Hu}}, \bibinfo {author} {\bibfnamefont {A.}~\bibnamefont
  {Stroppa}}, \ and\ \bibinfo {author} {\bibfnamefont {Q.}~\bibnamefont
  {Liu}},\ }\href {\doibase 10.1103/PhysRevLett.134.106802} {\bibfield
  {journal} {\bibinfo  {journal} {Phys. Rev. Lett.}\ }\textbf {\bibinfo
  {volume} {134}},\ \bibinfo {pages} {106802} (\bibinfo {year}
  {2025})}\BibitemShut {NoStop}%
\bibitem [{\citenamefont {Duan}\ \emph
  {et~al.}(2025{\natexlab{b}})\citenamefont {Duan}, \citenamefont {Zhang},
  \citenamefont {Zhu}, \citenamefont {Liu}, \citenamefont {Zhang},
  \citenamefont {\ifmmode \check{Z}\else \v{Z}\fi{}uti\ifmmode~\acute{c}\else
  \'{c}\fi{}},\ and\ \citenamefont {Zhou}}]{Duan2025Antiferroelectric}%
  \BibitemOpen
  \bibfield  {author} {\bibinfo {author} {\bibfnamefont {X.}~\bibnamefont
  {Duan}}, \bibinfo {author} {\bibfnamefont {J.}~\bibnamefont {Zhang}},
  \bibinfo {author} {\bibfnamefont {Z.}~\bibnamefont {Zhu}}, \bibinfo {author}
  {\bibfnamefont {Y.}~\bibnamefont {Liu}}, \bibinfo {author} {\bibfnamefont
  {Z.}~\bibnamefont {Zhang}}, \bibinfo {author} {\bibfnamefont
  {I.}~\bibnamefont {\ifmmode \check{Z}\else
  \v{Z}\fi{}uti\ifmmode~\acute{c}\else \'{c}\fi{}}}, \ and\ \bibinfo {author}
  {\bibfnamefont {T.}~\bibnamefont {Zhou}},\ }\href {\doibase
  10.1103/PhysRevLett.134.106801} {\bibfield  {journal} {\bibinfo  {journal}
  {Phys. Rev. Lett.}\ }\textbf {\bibinfo {volume} {134}},\ \bibinfo {pages}
  {106801} (\bibinfo {year} {2025}{\natexlab{b}})}\BibitemShut {NoStop}%
\bibitem [{\citenamefont {Zhu}\ \emph {et~al.}(2025{\natexlab{a}})\citenamefont
  {Zhu}, \citenamefont {Liu}, \citenamefont {Duan}, \citenamefont {Zhang},
  \citenamefont {Hao}, \citenamefont {Wei}, \citenamefont {Žutić},\ and\
  \citenamefont {Zhou}}]{Zhu2025Emergent}%
  \BibitemOpen
  \bibfield  {author} {\bibinfo {author} {\bibfnamefont {Z.}~\bibnamefont
  {Zhu}}, \bibinfo {author} {\bibfnamefont {Y.}~\bibnamefont {Liu}}, \bibinfo
  {author} {\bibfnamefont {X.}~\bibnamefont {Duan}}, \bibinfo {author}
  {\bibfnamefont {J.}~\bibnamefont {Zhang}}, \bibinfo {author} {\bibfnamefont
  {B.}~\bibnamefont {Hao}}, \bibinfo {author} {\bibfnamefont {S.-H.}\
  \bibnamefont {Wei}}, \bibinfo {author} {\bibfnamefont {I.}~\bibnamefont
  {Žutić}}, \ and\ \bibinfo {author} {\bibfnamefont {T.}~\bibnamefont
  {Zhou}},\ }\href {\doibase 10.1007/s11433-025-2778-3} {\bibfield  {journal}
  {\bibinfo  {journal} {Science China Physics, Mechanics amp; Astronomy}\
  }\textbf {\bibinfo {volume} {68}} (\bibinfo {year} {2025}{\natexlab{a}}),\
  10.1007/s11433-025-2778-3}\BibitemShut {NoStop}%
\bibitem [{\citenamefont {Zhu}\ \emph {et~al.}(2025{\natexlab{b}})\citenamefont
  {Zhu}, \citenamefont {Duan}, \citenamefont {Zhang}, \citenamefont {Hao},
  \citenamefont {Žutić},\ and\ \citenamefont {Zhou}}]{Zhu2025Twodim}%
  \BibitemOpen
  \bibfield  {author} {\bibinfo {author} {\bibfnamefont {Z.}~\bibnamefont
  {Zhu}}, \bibinfo {author} {\bibfnamefont {X.}~\bibnamefont {Duan}}, \bibinfo
  {author} {\bibfnamefont {J.}~\bibnamefont {Zhang}}, \bibinfo {author}
  {\bibfnamefont {B.}~\bibnamefont {Hao}}, \bibinfo {author} {\bibfnamefont
  {I.}~\bibnamefont {Žutić}}, \ and\ \bibinfo {author} {\bibfnamefont
  {T.}~\bibnamefont {Zhou}},\ }\href {\doibase 10.1021/acs.nanolett.5c02121}
  {\bibfield  {journal} {\bibinfo  {journal} {Nano Letters}\ }\textbf {\bibinfo
  {volume} {25}},\ \bibinfo {pages} {9456–9462} (\bibinfo {year}
  {2025}{\natexlab{b}})}\BibitemShut {NoStop}%
\bibitem [{\citenamefont {Denisov}\ and\ \citenamefont {\ifmmode \check{Z}\else
  \v{Z}\fi{}uti\ifmmode~\acute{c}\else
  \'{c}\fi{}}(2024)}]{denisov2024anisotropic}%
  \BibitemOpen
  \bibfield  {author} {\bibinfo {author} {\bibfnamefont {K.~S.}\ \bibnamefont
  {Denisov}}\ and\ \bibinfo {author} {\bibfnamefont {I.}~\bibnamefont {\ifmmode
  \check{Z}\else \v{Z}\fi{}uti\ifmmode~\acute{c}\else \'{c}\fi{}}},\ }\href
  {\doibase 10.1103/PhysRevB.110.L180403} {\bibfield  {journal} {\bibinfo
  {journal} {Phys. Rev. B}\ }\textbf {\bibinfo {volume} {110}},\ \bibinfo
  {pages} {L180403} (\bibinfo {year} {2024})}\BibitemShut {NoStop}%
\bibitem [{\citenamefont {Cao}\ and\ \citenamefont
  {Žutić}(2025)}]{Cao2025Symmetry}%
  \BibitemOpen
  \bibfield  {author} {\bibinfo {author} {\bibfnamefont {D.~K. S. L.~Y.}\
  \bibnamefont {Cao}, \bibfnamefont {Jiayu~David}}\ and\ \bibinfo {author}
  {\bibfnamefont {I.}~\bibnamefont {Žutić}},\ }\href {\doibase
  10.1103/zn7r-k1xd} {\bibfield  {journal} {\bibinfo  {journal} {Phys. Rev.
  Lett.}\ ,\ } (\bibinfo {year} {2025})}\BibitemShut {NoStop}%
\bibitem [{\citenamefont {Mazin}(2022{\natexlab{b}})}]{mazin2022notes}%
  \BibitemOpen
  \bibfield  {author} {\bibinfo {author} {\bibfnamefont {I.~I.}\ \bibnamefont
  {Mazin}},\ }\href {https://arxiv.org/abs/2203.05000} {\bibfield  {journal}
  {\bibinfo  {journal} {arXiv preprint arXiv:2203.05000}\ } (\bibinfo {year}
  {2022}{\natexlab{b}})}\BibitemShut {NoStop}%
\bibitem [{\citenamefont {Casola}\ \emph {et~al.}(2018)\citenamefont {Casola},
  \citenamefont {Van Der~Sar},\ and\ \citenamefont
  {Yacoby}}]{casola2018probing}%
  \BibitemOpen
  \bibfield  {author} {\bibinfo {author} {\bibfnamefont {F.}~\bibnamefont
  {Casola}}, \bibinfo {author} {\bibfnamefont {T.}~\bibnamefont {Van Der~Sar}},
  \ and\ \bibinfo {author} {\bibfnamefont {A.}~\bibnamefont {Yacoby}},\ }\href
  {https://www.nature.com/articles/natrevmats201788} {\bibfield  {journal}
  {\bibinfo  {journal} {Nature Reviews Materials}\ }\textbf {\bibinfo {volume}
  {3}},\ \bibinfo {pages} {1} (\bibinfo {year} {2018})}\BibitemShut {NoStop}%
\bibitem [{\citenamefont {Hong}\ \emph {et~al.}(2013)\citenamefont {Hong},
  \citenamefont {Grinolds}, \citenamefont {Pham}, \citenamefont {Le~Sage},
  \citenamefont {Luan}, \citenamefont {Walsworth},\ and\ \citenamefont
  {Yacoby}}]{hong2013nanoscale}%
  \BibitemOpen
  \bibfield  {author} {\bibinfo {author} {\bibfnamefont {S.}~\bibnamefont
  {Hong}}, \bibinfo {author} {\bibfnamefont {M.~S.}\ \bibnamefont {Grinolds}},
  \bibinfo {author} {\bibfnamefont {L.~M.}\ \bibnamefont {Pham}}, \bibinfo
  {author} {\bibfnamefont {D.}~\bibnamefont {Le~Sage}}, \bibinfo {author}
  {\bibfnamefont {L.}~\bibnamefont {Luan}}, \bibinfo {author} {\bibfnamefont
  {R.~L.}\ \bibnamefont {Walsworth}}, \ and\ \bibinfo {author} {\bibfnamefont
  {A.}~\bibnamefont {Yacoby}},\ }\href
  {https://www.cambridge.org/core/journals/mrs-bulletin/article/abs/nanoscale-magnetometry-with-nv-centers-in-diamond/0EAD9D71BB13F6B6508CE68023DBEEA2}
  {\bibfield  {journal} {\bibinfo  {journal} {MRS bulletin}\ }\textbf {\bibinfo
  {volume} {38}},\ \bibinfo {pages} {155} (\bibinfo {year} {2013})}\BibitemShut
  {NoStop}%
\bibitem [{\citenamefont {Kolkowitz}\ \emph {et~al.}(2012)\citenamefont
  {Kolkowitz}, \citenamefont {Unterreithmeier}, \citenamefont {Bennett},\ and\
  \citenamefont {Lukin}}]{kolkowitz2012sensing}%
  \BibitemOpen
  \bibfield  {author} {\bibinfo {author} {\bibfnamefont {S.}~\bibnamefont
  {Kolkowitz}}, \bibinfo {author} {\bibfnamefont {Q.~P.}\ \bibnamefont
  {Unterreithmeier}}, \bibinfo {author} {\bibfnamefont {S.~D.}\ \bibnamefont
  {Bennett}}, \ and\ \bibinfo {author} {\bibfnamefont {M.~D.}\ \bibnamefont
  {Lukin}},\ }\href
  {https://journals.aps.org/prl/abstract/10.1103/PhysRevLett.109.137601}
  {\bibfield  {journal} {\bibinfo  {journal} {Phys. Rev. Lett.}\ }\textbf
  {\bibinfo {volume} {109}},\ \bibinfo {pages} {137601} (\bibinfo {year}
  {2012})}\BibitemShut {NoStop}%
\bibitem [{\citenamefont {Dovzhenko}\ \emph {et~al.}(2018)\citenamefont
  {Dovzhenko}, \citenamefont {Casola}, \citenamefont {Schlotter}, \citenamefont
  {Zhou}, \citenamefont {B{\"u}ttner}, \citenamefont {Walsworth}, \citenamefont
  {Beach},\ and\ \citenamefont {Yacoby}}]{dovzhenko2018magnetostatic}%
  \BibitemOpen
  \bibfield  {author} {\bibinfo {author} {\bibfnamefont {Y.}~\bibnamefont
  {Dovzhenko}}, \bibinfo {author} {\bibfnamefont {F.}~\bibnamefont {Casola}},
  \bibinfo {author} {\bibfnamefont {S.}~\bibnamefont {Schlotter}}, \bibinfo
  {author} {\bibfnamefont {T.}~\bibnamefont {Zhou}}, \bibinfo {author}
  {\bibfnamefont {F.}~\bibnamefont {B{\"u}ttner}}, \bibinfo {author}
  {\bibfnamefont {R.}~\bibnamefont {Walsworth}}, \bibinfo {author}
  {\bibfnamefont {G.}~\bibnamefont {Beach}}, \ and\ \bibinfo {author}
  {\bibfnamefont {A.}~\bibnamefont {Yacoby}},\ }\href
  {https://www.nature.com/articles/s41467-018-05158-9} {\bibfield  {journal}
  {\bibinfo  {journal} {Nature Communications}\ }\textbf {\bibinfo {volume}
  {9}},\ \bibinfo {pages} {1} (\bibinfo {year} {2018})}\BibitemShut {NoStop}%
\bibitem [{\citenamefont {Hsieh}\ \emph {et~al.}(2019)\citenamefont {Hsieh},
  \citenamefont {Bhattacharyya}, \citenamefont {Zu}, \citenamefont {Mittiga},
  \citenamefont {Smart}, \citenamefont {Machado}, \citenamefont {Kobrin},
  \citenamefont {H{\"o}hn}, \citenamefont {Rui}, \citenamefont {Kamrani} \emph
  {et~al.}}]{hsieh2019imaging}%
  \BibitemOpen
  \bibfield  {author} {\bibinfo {author} {\bibfnamefont {S.}~\bibnamefont
  {Hsieh}}, \bibinfo {author} {\bibfnamefont {P.}~\bibnamefont
  {Bhattacharyya}}, \bibinfo {author} {\bibfnamefont {C.}~\bibnamefont {Zu}},
  \bibinfo {author} {\bibfnamefont {T.}~\bibnamefont {Mittiga}}, \bibinfo
  {author} {\bibfnamefont {T.}~\bibnamefont {Smart}}, \bibinfo {author}
  {\bibfnamefont {F.}~\bibnamefont {Machado}}, \bibinfo {author} {\bibfnamefont
  {B.}~\bibnamefont {Kobrin}}, \bibinfo {author} {\bibfnamefont
  {T.}~\bibnamefont {H{\"o}hn}}, \bibinfo {author} {\bibfnamefont
  {N.}~\bibnamefont {Rui}}, \bibinfo {author} {\bibfnamefont {M.}~\bibnamefont
  {Kamrani}},  \emph {et~al.},\ }\href
  {https://www.science.org/doi/10.1126/science.aaw4352} {\bibfield  {journal}
  {\bibinfo  {journal} {Science}\ }\textbf {\bibinfo {volume} {366}},\ \bibinfo
  {pages} {1349} (\bibinfo {year} {2019})}\BibitemShut {NoStop}%
\bibitem [{\citenamefont {Rovny}\ \emph {et~al.}(2024)\citenamefont {Rovny},
  \citenamefont {Gopalakrishnan}, \citenamefont {Jayich}, \citenamefont
  {Maletinsky}, \citenamefont {Demler},\ and\ \citenamefont
  {de~Leon}}]{rovny2024new}%
  \BibitemOpen
  \bibfield  {author} {\bibinfo {author} {\bibfnamefont {J.}~\bibnamefont
  {Rovny}}, \bibinfo {author} {\bibfnamefont {S.}~\bibnamefont
  {Gopalakrishnan}}, \bibinfo {author} {\bibfnamefont {A.~C.~B.}\ \bibnamefont
  {Jayich}}, \bibinfo {author} {\bibfnamefont {P.}~\bibnamefont {Maletinsky}},
  \bibinfo {author} {\bibfnamefont {E.}~\bibnamefont {Demler}}, \ and\ \bibinfo
  {author} {\bibfnamefont {N.~P.}\ \bibnamefont {de~Leon}},\ }\href
  {https://arxiv.org/abs/2403.13710} {\bibfield  {journal} {\bibinfo  {journal}
  {arXiv preprint arXiv:2403.13710}\ } (\bibinfo {year} {2024})}\BibitemShut
  {NoStop}%
\bibitem [{\citenamefont {Du}\ \emph {et~al.}(2017)\citenamefont {Du},
  \citenamefont {van~der Sar}, \citenamefont {Zhou}, \citenamefont {Upadhyaya},
  \citenamefont {Casola}, \citenamefont {Zhang}, \citenamefont {Onbasli},
  \citenamefont {Ross}, \citenamefont {Walsworth}, \citenamefont
  {Tserkovnyak},\ and\ \citenamefont
  {Yacoby}}]{du2017controlandlocalmeasurement}%
  \BibitemOpen
  \bibfield  {author} {\bibinfo {author} {\bibfnamefont {C.}~\bibnamefont
  {Du}}, \bibinfo {author} {\bibfnamefont {T.}~\bibnamefont {van~der Sar}},
  \bibinfo {author} {\bibfnamefont {T.~X.}\ \bibnamefont {Zhou}}, \bibinfo
  {author} {\bibfnamefont {P.}~\bibnamefont {Upadhyaya}}, \bibinfo {author}
  {\bibfnamefont {F.}~\bibnamefont {Casola}}, \bibinfo {author} {\bibfnamefont
  {H.}~\bibnamefont {Zhang}}, \bibinfo {author} {\bibfnamefont {M.~C.}\
  \bibnamefont {Onbasli}}, \bibinfo {author} {\bibfnamefont {C.~A.}\
  \bibnamefont {Ross}}, \bibinfo {author} {\bibfnamefont {R.~L.}\ \bibnamefont
  {Walsworth}}, \bibinfo {author} {\bibfnamefont {Y.}~\bibnamefont
  {Tserkovnyak}}, \ and\ \bibinfo {author} {\bibfnamefont {A.}~\bibnamefont
  {Yacoby}},\ }\href {\doibase 10.1126/science.aak9611} {\bibfield  {journal}
  {\bibinfo  {journal} {Science}\ }\textbf {\bibinfo {volume} {357}},\ \bibinfo
  {pages} {195} (\bibinfo {year} {2017})}\BibitemShut {NoStop}%
\bibitem [{\citenamefont {Wang}\ \emph {et~al.}(2022)\citenamefont {Wang},
  \citenamefont {Zhang}, \citenamefont {McLaughlin}, \citenamefont {Flebus},
  \citenamefont {Huang}, \citenamefont {Xiao}, \citenamefont {Liu},
  \citenamefont {Wu}, \citenamefont {Fullerton}, \citenamefont {Tserkovnyak},\
  and\ \citenamefont {Du}}]{wang_2022_noninvasive_measurements}%
  \BibitemOpen
  \bibfield  {author} {\bibinfo {author} {\bibfnamefont {H.}~\bibnamefont
  {Wang}}, \bibinfo {author} {\bibfnamefont {S.}~\bibnamefont {Zhang}},
  \bibinfo {author} {\bibfnamefont {N.~J.}\ \bibnamefont {McLaughlin}},
  \bibinfo {author} {\bibfnamefont {B.}~\bibnamefont {Flebus}}, \bibinfo
  {author} {\bibfnamefont {M.}~\bibnamefont {Huang}}, \bibinfo {author}
  {\bibfnamefont {Y.}~\bibnamefont {Xiao}}, \bibinfo {author} {\bibfnamefont
  {C.}~\bibnamefont {Liu}}, \bibinfo {author} {\bibfnamefont {M.}~\bibnamefont
  {Wu}}, \bibinfo {author} {\bibfnamefont {E.~E.}\ \bibnamefont {Fullerton}},
  \bibinfo {author} {\bibfnamefont {Y.}~\bibnamefont {Tserkovnyak}}, \ and\
  \bibinfo {author} {\bibfnamefont {C.~R.}\ \bibnamefont {Du}},\ }\href
  {\doibase 10.1126/sciadv.abg8562} {\bibfield  {journal} {\bibinfo  {journal}
  {Science Advances}\ }\textbf {\bibinfo {volume} {8}},\ \bibinfo {pages}
  {eabg8562} (\bibinfo {year} {2022})}\BibitemShut {NoStop}%
\bibitem [{\citenamefont {Heitzer}\ \emph {et~al.}(2024)\citenamefont
  {Heitzer}, \citenamefont {Pinto}, \citenamefont {Rodr\'{\i}guez},
  \citenamefont {Rodr\'{\i}guez-Su\'arez},\ and\ \citenamefont
  {Maze}}]{Heitzer_2024_Characterization}%
  \BibitemOpen
  \bibfield  {author} {\bibinfo {author} {\bibfnamefont {R.~C.}\ \bibnamefont
  {Heitzer}}, \bibinfo {author} {\bibfnamefont {F.}~\bibnamefont {Pinto}},
  \bibinfo {author} {\bibfnamefont {E.}~\bibnamefont {Rodr\'{\i}guez}},
  \bibinfo {author} {\bibfnamefont {R.}~\bibnamefont
  {Rodr\'{\i}guez-Su\'arez}}, \ and\ \bibinfo {author} {\bibfnamefont {J.~R.}\
  \bibnamefont {Maze}},\ }\href {\doibase 10.1103/PhysRevB.110.134431}
  {\bibfield  {journal} {\bibinfo  {journal} {Phys. Rev. B}\ }\textbf {\bibinfo
  {volume} {110}},\ \bibinfo {pages} {134431} (\bibinfo {year}
  {2024})}\BibitemShut {NoStop}%
\bibitem [{\citenamefont {Flebus}\ and\ \citenamefont
  {Tserkovnyak}(2018)}]{flebus_2018_quantum_impurity_relaxometry}%
  \BibitemOpen
  \bibfield  {author} {\bibinfo {author} {\bibfnamefont {B.}~\bibnamefont
  {Flebus}}\ and\ \bibinfo {author} {\bibfnamefont {Y.}~\bibnamefont
  {Tserkovnyak}},\ }\href {\doibase 10.1103/PhysRevLett.121.187204} {\bibfield
  {journal} {\bibinfo  {journal} {Phys. Rev. Lett.}\ }\textbf {\bibinfo
  {volume} {121}},\ \bibinfo {pages} {187204} (\bibinfo {year}
  {2018})}\BibitemShut {NoStop}%
\bibitem [{\citenamefont {Chatterjee}\ \emph {et~al.}(2019)\citenamefont
  {Chatterjee}, \citenamefont {Rodriguez-Nieva},\ and\ \citenamefont
  {Demler}}]{chatterjee2019diagnosingphases}%
  \BibitemOpen
  \bibfield  {author} {\bibinfo {author} {\bibfnamefont {S.}~\bibnamefont
  {Chatterjee}}, \bibinfo {author} {\bibfnamefont {J.~F.}\ \bibnamefont
  {Rodriguez-Nieva}}, \ and\ \bibinfo {author} {\bibfnamefont {E.}~\bibnamefont
  {Demler}},\ }\href {\doibase 10.1103/PhysRevB.99.104425} {\bibfield
  {journal} {\bibinfo  {journal} {Phys. Rev. B}\ }\textbf {\bibinfo {volume}
  {99}},\ \bibinfo {pages} {104425} (\bibinfo {year} {2019})}\BibitemShut
  {NoStop}%
\bibitem [{\citenamefont {Langsjoen}\ \emph {et~al.}(2012)\citenamefont
  {Langsjoen}, \citenamefont {Poudel}, \citenamefont {Vavilov},\ and\
  \citenamefont {Joynt}}]{langsjoen2012qubit}%
  \BibitemOpen
  \bibfield  {author} {\bibinfo {author} {\bibfnamefont {L.~S.}\ \bibnamefont
  {Langsjoen}}, \bibinfo {author} {\bibfnamefont {A.}~\bibnamefont {Poudel}},
  \bibinfo {author} {\bibfnamefont {M.~G.}\ \bibnamefont {Vavilov}}, \ and\
  \bibinfo {author} {\bibfnamefont {R.}~\bibnamefont {Joynt}},\ }\href
  {https://journals.aps.org/pra/abstract/10.1103/PhysRevA.86.010301} {\bibfield
   {journal} {\bibinfo  {journal} {Phys. Rev. A}\ }\textbf {\bibinfo {volume}
  {86}},\ \bibinfo {pages} {010301} (\bibinfo {year} {2012})}\BibitemShut
  {NoStop}%
\bibitem [{\citenamefont {Agarwal}\ \emph {et~al.}(2017)\citenamefont
  {Agarwal}, \citenamefont {Schmidt}, \citenamefont {Halperin}, \citenamefont
  {Oganesyan}, \citenamefont {Zar\'and}, \citenamefont {Lukin},\ and\
  \citenamefont {Demler}}]{Agarwal_currentNoise2017}%
  \BibitemOpen
  \bibfield  {author} {\bibinfo {author} {\bibfnamefont {K.}~\bibnamefont
  {Agarwal}}, \bibinfo {author} {\bibfnamefont {R.}~\bibnamefont {Schmidt}},
  \bibinfo {author} {\bibfnamefont {B.}~\bibnamefont {Halperin}}, \bibinfo
  {author} {\bibfnamefont {V.}~\bibnamefont {Oganesyan}}, \bibinfo {author}
  {\bibfnamefont {G.}~\bibnamefont {Zar\'and}}, \bibinfo {author}
  {\bibfnamefont {M.~D.}\ \bibnamefont {Lukin}}, \ and\ \bibinfo {author}
  {\bibfnamefont {E.}~\bibnamefont {Demler}},\ }\href {\doibase
  10.1103/PhysRevB.95.155107} {\bibfield  {journal} {\bibinfo  {journal} {Phys.
  Rev. B}\ }\textbf {\bibinfo {volume} {95}},\ \bibinfo {pages} {155107}
  (\bibinfo {year} {2017})}\BibitemShut {NoStop}%
\bibitem [{\citenamefont {Rodriguez-Nieva}\ \emph {et~al.}(2018)\citenamefont
  {Rodriguez-Nieva}, \citenamefont {Agarwal}, \citenamefont {Giamarchi},
  \citenamefont {Halperin}, \citenamefont {Lukin},\ and\ \citenamefont
  {Demler}}]{Rodriguez-Nieva_1D_2018}%
  \BibitemOpen
  \bibfield  {author} {\bibinfo {author} {\bibfnamefont {J.~F.}\ \bibnamefont
  {Rodriguez-Nieva}}, \bibinfo {author} {\bibfnamefont {K.}~\bibnamefont
  {Agarwal}}, \bibinfo {author} {\bibfnamefont {T.}~\bibnamefont {Giamarchi}},
  \bibinfo {author} {\bibfnamefont {B.~I.}\ \bibnamefont {Halperin}}, \bibinfo
  {author} {\bibfnamefont {M.~D.}\ \bibnamefont {Lukin}}, \ and\ \bibinfo
  {author} {\bibfnamefont {E.}~\bibnamefont {Demler}},\ }\href {\doibase
  10.1103/PhysRevB.98.195433} {\bibfield  {journal} {\bibinfo  {journal} {Phys.
  Rev. B}\ }\textbf {\bibinfo {volume} {98}},\ \bibinfo {pages} {195433}
  (\bibinfo {year} {2018})}\BibitemShut {NoStop}%
\bibitem [{\citenamefont {Rodriguez-Nieva}\ \emph {et~al.}(2022)\citenamefont
  {Rodriguez-Nieva}, \citenamefont {Podolsky},\ and\ \citenamefont
  {Demler}}]{rodriguez2022probing}%
  \BibitemOpen
  \bibfield  {author} {\bibinfo {author} {\bibfnamefont {J.~F.}\ \bibnamefont
  {Rodriguez-Nieva}}, \bibinfo {author} {\bibfnamefont {D.}~\bibnamefont
  {Podolsky}}, \ and\ \bibinfo {author} {\bibfnamefont {E.}~\bibnamefont
  {Demler}},\ }\href
  {https://journals.aps.org/prb/abstract/10.1103/PhysRevB.105.174412}
  {\bibfield  {journal} {\bibinfo  {journal} {Phys. Rev. B}\ }\textbf {\bibinfo
  {volume} {105}},\ \bibinfo {pages} {174412} (\bibinfo {year}
  {2022})}\BibitemShut {NoStop}%
\bibitem [{\citenamefont {\ifmmode~\check{S}\else \v{S}\fi{}mejkal}\ \emph
  {et~al.}(2023)\citenamefont {\ifmmode~\check{S}\else \v{S}\fi{}mejkal},
  \citenamefont {Marmodoro}, \citenamefont {Ahn}, \citenamefont
  {Gonz\'alez-Hern\'andez}, \citenamefont {Turek}, \citenamefont {Mankovsky},
  \citenamefont {Ebert}, \citenamefont {D'Souza}, \citenamefont
  {\ifmmode~\check{S}\else \v{S}\fi{}ipr}, \citenamefont {Sinova},\ and\
  \citenamefont {Jungwirth}}]{smejkal_2023_chiral_magnons}%
  \BibitemOpen
  \bibfield  {author} {\bibinfo {author} {\bibfnamefont {L.}~\bibnamefont
  {\ifmmode~\check{S}\else \v{S}\fi{}mejkal}}, \bibinfo {author} {\bibfnamefont
  {A.}~\bibnamefont {Marmodoro}}, \bibinfo {author} {\bibfnamefont {K.-H.}\
  \bibnamefont {Ahn}}, \bibinfo {author} {\bibfnamefont {R.}~\bibnamefont
  {Gonz\'alez-Hern\'andez}}, \bibinfo {author} {\bibfnamefont {I.}~\bibnamefont
  {Turek}}, \bibinfo {author} {\bibfnamefont {S.}~\bibnamefont {Mankovsky}},
  \bibinfo {author} {\bibfnamefont {H.}~\bibnamefont {Ebert}}, \bibinfo
  {author} {\bibfnamefont {S.~W.}\ \bibnamefont {D'Souza}}, \bibinfo {author}
  {\bibfnamefont {O.}~\bibnamefont {\ifmmode~\check{S}\else \v{S}\fi{}ipr}},
  \bibinfo {author} {\bibfnamefont {J.}~\bibnamefont {Sinova}}, \ and\ \bibinfo
  {author} {\bibfnamefont {T.}~\bibnamefont {Jungwirth}},\ }\href {\doibase
  10.1103/PhysRevLett.131.256703} {\bibfield  {journal} {\bibinfo  {journal}
  {Phys. Rev. Lett.}\ }\textbf {\bibinfo {volume} {131}},\ \bibinfo {pages}
  {256703} (\bibinfo {year} {2023})}\BibitemShut {NoStop}%
\bibitem [{\citenamefont {Mzyk}\ \emph {et~al.}(2022)\citenamefont {Mzyk},
  \citenamefont {Sigaeva},\ and\ \citenamefont
  {Schirhagl}}]{mzyk2022relaxometry}%
  \BibitemOpen
  \bibfield  {author} {\bibinfo {author} {\bibfnamefont {A.}~\bibnamefont
  {Mzyk}}, \bibinfo {author} {\bibfnamefont {A.}~\bibnamefont {Sigaeva}}, \
  and\ \bibinfo {author} {\bibfnamefont {R.}~\bibnamefont {Schirhagl}},\ }\href
  {\doibase 10.1021/acs.accounts.2c00520} {\bibfield  {journal} {\bibinfo
  {journal} {Accounts of Chemical Research}\ }\textbf {\bibinfo {volume}
  {55}},\ \bibinfo {pages} {3572} (\bibinfo {year} {2022})}\BibitemShut
  {NoStop}%
\bibitem [{\citenamefont {\ifmmode~\check{S}\else \v{S}\fi{}mejkal}\ \emph
  {et~al.}(2022{\natexlab{d}})\citenamefont {\ifmmode~\check{S}\else
  \v{S}\fi{}mejkal}, \citenamefont {Sinova},\ and\ \citenamefont
  {Jungwirth}}]{smejkal_2022_beyond_conventional}%
  \BibitemOpen
  \bibfield  {author} {\bibinfo {author} {\bibfnamefont {L.}~\bibnamefont
  {\ifmmode~\check{S}\else \v{S}\fi{}mejkal}}, \bibinfo {author} {\bibfnamefont
  {J.}~\bibnamefont {Sinova}}, \ and\ \bibinfo {author} {\bibfnamefont
  {T.}~\bibnamefont {Jungwirth}},\ }\href {\doibase 10.1103/PhysRevX.12.031042}
  {\bibfield  {journal} {\bibinfo  {journal} {Phys. Rev. X}\ }\textbf {\bibinfo
  {volume} {12}},\ \bibinfo {pages} {031042} (\bibinfo {year}
  {2022}{\natexlab{d}})}\BibitemShut {NoStop}%
\bibitem [{\citenamefont {\ifmmode~\check{S}\else \v{S}\fi{}mejkal}\ \emph
  {et~al.}(2022{\natexlab{e}})\citenamefont {\ifmmode~\check{S}\else
  \v{S}\fi{}mejkal}, \citenamefont {Sinova},\ and\ \citenamefont
  {Jungwirth}}]{smejkal_2022_emerging_research}%
  \BibitemOpen
  \bibfield  {author} {\bibinfo {author} {\bibfnamefont {L.}~\bibnamefont
  {\ifmmode~\check{S}\else \v{S}\fi{}mejkal}}, \bibinfo {author} {\bibfnamefont
  {J.}~\bibnamefont {Sinova}}, \ and\ \bibinfo {author} {\bibfnamefont
  {T.}~\bibnamefont {Jungwirth}},\ }\href {\doibase 10.1103/PhysRevX.12.040501}
  {\bibfield  {journal} {\bibinfo  {journal} {Phys. Rev. X}\ }\textbf {\bibinfo
  {volume} {12}},\ \bibinfo {pages} {040501} (\bibinfo {year}
  {2022}{\natexlab{e}})}\BibitemShut {NoStop}%
\bibitem [{\citenamefont {Eto}\ \emph {et~al.}(2025)\citenamefont {Eto},
  \citenamefont {Gohlke}, \citenamefont {Sinova}, \citenamefont {Mochizuki},
  \citenamefont {Chernyshev},\ and\ \citenamefont
  {Mook}}]{eto2025spontaneousmagnon}%
  \BibitemOpen
  \bibfield  {author} {\bibinfo {author} {\bibfnamefont {R.}~\bibnamefont
  {Eto}}, \bibinfo {author} {\bibfnamefont {M.}~\bibnamefont {Gohlke}},
  \bibinfo {author} {\bibfnamefont {J.}~\bibnamefont {Sinova}}, \bibinfo
  {author} {\bibfnamefont {M.}~\bibnamefont {Mochizuki}}, \bibinfo {author}
  {\bibfnamefont {A.~L.}\ \bibnamefont {Chernyshev}}, \ and\ \bibinfo {author}
  {\bibfnamefont {A.}~\bibnamefont {Mook}},\ }\href
  {https://arxiv.org/abs/2502.20146} {\bibfield  {journal} {\bibinfo  {journal}
  {arXiv[cond-mat.str-el]:}\ }\textbf {\bibinfo {volume} {2502.20146}}
  (\bibinfo {year} {2025})}\BibitemShut {NoStop}%
\bibitem [{\citenamefont {Kubo}(1966)}]{kubo_1966_fluctuation_dissipation}%
  \BibitemOpen
  \bibfield  {author} {\bibinfo {author} {\bibfnamefont {R.}~\bibnamefont
  {Kubo}},\ }\href
  {https://iopscience.iop.org/article/10.1088/0034-4885/29/1/306/meta}
  {\bibfield  {journal} {\bibinfo  {journal} {Reports on Progress in Physics}\
  }\textbf {\bibinfo {volume} {29}} (\bibinfo {year} {1966})}\BibitemShut
  {NoStop}%
\bibitem [{\citenamefont {Guslienko}\ and\ \citenamefont
  {Slavin}(2011)}]{guslienki2011magnetostaticgreensfunction}%
  \BibitemOpen
  \bibfield  {author} {\bibinfo {author} {\bibfnamefont {K.~Y.}\ \bibnamefont
  {Guslienko}}\ and\ \bibinfo {author} {\bibfnamefont {A.~N.}\ \bibnamefont
  {Slavin}},\ }\href {\doibase https://doi.org/10.1016/j.jmmm.2011.05.020}
  {\bibfield  {journal} {\bibinfo  {journal} {Journal of Magnetism and Magnetic
  Materials}\ }\textbf {\bibinfo {volume} {323}},\ \bibinfo {pages} {2418}
  (\bibinfo {year} {2011})}\BibitemShut {NoStop}%
\bibitem [{\citenamefont {Rezende}\ \emph {et~al.}(2016)\citenamefont
  {Rezende}, \citenamefont {Rodr\'{i}guez-Su\'{a}rez}, \citenamefont {Cunha},
  \citenamefont {{L\'{o}pez Ortiz}},\ and\ \citenamefont
  {Azevedo}}]{rezende_2016_bulk_magnon}%
  \BibitemOpen
  \bibfield  {author} {\bibinfo {author} {\bibfnamefont {S.}~\bibnamefont
  {Rezende}}, \bibinfo {author} {\bibfnamefont {R.}~\bibnamefont
  {Rodr\'{i}guez-Su\'{a}rez}}, \bibinfo {author} {\bibfnamefont
  {R.}~\bibnamefont {Cunha}}, \bibinfo {author} {\bibfnamefont
  {J.}~\bibnamefont {{L\'{o}pez Ortiz}}}, \ and\ \bibinfo {author}
  {\bibfnamefont {A.}~\bibnamefont {Azevedo}},\ }\href {\doibase
  https://doi.org/10.1016/j.jmmm.2015.07.102} {\bibfield  {journal} {\bibinfo
  {journal} {Journal of Magnetism and Magnetic Materials}\ }\textbf {\bibinfo
  {volume} {400}},\ \bibinfo {pages} {171} (\bibinfo {year} {2016})},\ \bibinfo
  {note} {proceedings of the 20th International Conference on Magnetism
  (Barcelona) 5-10 July 2015}\BibitemShut {NoStop}%
\bibitem [{\citenamefont {Rezende}\ \emph {et~al.}(2019)\citenamefont
  {Rezende}, \citenamefont {Azevedo},\ and\ \citenamefont
  {Rodr\'{i}guez-Su\'{a}rez}}]{rezende_2019_introduction}%
  \BibitemOpen
  \bibfield  {author} {\bibinfo {author} {\bibfnamefont {S.~M.}\ \bibnamefont
  {Rezende}}, \bibinfo {author} {\bibfnamefont {A.}~\bibnamefont {Azevedo}}, \
  and\ \bibinfo {author} {\bibfnamefont {R.~L.}\ \bibnamefont
  {Rodr\'{i}guez-Su\'{a}rez}},\ }\href {\doibase 10.1063/1.5109132} {\bibfield
  {journal} {\bibinfo  {journal} {Journal of Applied Physics}\ }\textbf
  {\bibinfo {volume} {126}},\ \bibinfo {pages} {151101} (\bibinfo {year}
  {2019})}\BibitemShut {NoStop}%
\bibitem [{SM()}]{SM}%
  \BibitemOpen
  \href@noop {} {}\bibinfo {note} {See Supplementary Material for detailed
  calculations of the magnon band structure, the diffusion response function,
  and explicit expressions for the relaxation rate, which includes Refs.
  [77,78].}\BibitemShut {Stop}%
\bibitem [{\citenamefont {Brekke}\ \emph {et~al.}(2023)\citenamefont {Brekke},
  \citenamefont {Brataas},\ and\ \citenamefont
  {Sudb\o{}}}]{PhysRevB.108.224421}%
  \BibitemOpen
  \bibfield  {author} {\bibinfo {author} {\bibfnamefont {B.}~\bibnamefont
  {Brekke}}, \bibinfo {author} {\bibfnamefont {A.}~\bibnamefont {Brataas}}, \
  and\ \bibinfo {author} {\bibfnamefont {A.}~\bibnamefont {Sudb\o{}}},\ }\href
  {\doibase 10.1103/PhysRevB.108.224421} {\bibfield  {journal} {\bibinfo
  {journal} {Phys. Rev. B}\ }\textbf {\bibinfo {volume} {108}},\ \bibinfo
  {pages} {224421} (\bibinfo {year} {2023})}\BibitemShut {NoStop}%
\bibitem [{\citenamefont {Cui}\ \emph {et~al.}(2023)\citenamefont {Cui},
  \citenamefont {Zeng}, \citenamefont {Cui}, \citenamefont {Yu},\ and\
  \citenamefont {Yang}}]{cui_2023_efficient_spin}%
  \BibitemOpen
  \bibfield  {author} {\bibinfo {author} {\bibfnamefont {Q.}~\bibnamefont
  {Cui}}, \bibinfo {author} {\bibfnamefont {B.}~\bibnamefont {Zeng}}, \bibinfo
  {author} {\bibfnamefont {P.}~\bibnamefont {Cui}}, \bibinfo {author}
  {\bibfnamefont {T.}~\bibnamefont {Yu}}, \ and\ \bibinfo {author}
  {\bibfnamefont {H.}~\bibnamefont {Yang}},\ }\href {\doibase
  10.1103/PhysRevB.108.L180401} {\bibfield  {journal} {\bibinfo  {journal}
  {Phys. Rev. B}\ }\textbf {\bibinfo {volume} {108}},\ \bibinfo {pages}
  {L180401} (\bibinfo {year} {2023})}\BibitemShut {NoStop}%
\bibitem [{\citenamefont {Das}\ \emph {et~al.}(2024)\citenamefont {Das},
  \citenamefont {Leeb}, \citenamefont {Knolle},\ and\ \citenamefont
  {Knap}}]{das_2024_realizing_altermagnetism}%
  \BibitemOpen
  \bibfield  {author} {\bibinfo {author} {\bibfnamefont {P.}~\bibnamefont
  {Das}}, \bibinfo {author} {\bibfnamefont {V.}~\bibnamefont {Leeb}}, \bibinfo
  {author} {\bibfnamefont {J.}~\bibnamefont {Knolle}}, \ and\ \bibinfo {author}
  {\bibfnamefont {M.}~\bibnamefont {Knap}},\ }\href {\doibase
  10.1103/PhysRevLett.132.263402} {\bibfield  {journal} {\bibinfo  {journal}
  {Phys. Rev. Lett.}\ }\textbf {\bibinfo {volume} {132}},\ \bibinfo {pages}
  {263402} (\bibinfo {year} {2024})}\BibitemShut {NoStop}%
\bibitem [{\citenamefont {González}\ \emph {et~al.}(2025)\citenamefont
  {González}, \citenamefont {León}, \citenamefont {González-Fuentes},\ and\
  \citenamefont {Gallardo}}]{D4NR04053H}%
  \BibitemOpen
  \bibfield  {author} {\bibinfo {author} {\bibfnamefont {J.~W.}\ \bibnamefont
  {González}}, \bibinfo {author} {\bibfnamefont {A.~M.}\ \bibnamefont
  {León}}, \bibinfo {author} {\bibfnamefont {C.}~\bibnamefont
  {González-Fuentes}}, \ and\ \bibinfo {author} {\bibfnamefont {R.~A.}\
  \bibnamefont {Gallardo}},\ }\href {\doibase 10.1039/D4NR04053H} {\bibfield
  {journal} {\bibinfo  {journal} {Nanoscale}\ }\textbf {\bibinfo {volume}
  {17}},\ \bibinfo {pages} {4796} (\bibinfo {year} {2025})}\BibitemShut
  {NoStop}%
\bibitem [{\citenamefont {Lorenzo~Lanzini}(2025)}]{Lanzini2025Dual}%
  \BibitemOpen
  \bibfield  {author} {\bibinfo {author} {\bibfnamefont {M.~K.}\ \bibnamefont
  {Lorenzo~Lanzini}, \bibfnamefont {Purnendu~Das}},\ }\href
  {https://arxiv.org/abs/2506.03263} {\bibfield  {journal} {\bibinfo  {journal}
  {arXiv preprint arXiv:2506.03263}\ } (\bibinfo {year} {2025})}\BibitemShut
  {NoStop}%
\bibitem [{\citenamefont {Corticelli}\ \emph {et~al.}(2022)\citenamefont
  {Corticelli}, \citenamefont {Moessner},\ and\ \citenamefont
  {McClarty}}]{corticelli2022spinspacegroups}%
  \BibitemOpen
  \bibfield  {author} {\bibinfo {author} {\bibfnamefont {A.}~\bibnamefont
  {Corticelli}}, \bibinfo {author} {\bibfnamefont {R.}~\bibnamefont
  {Moessner}}, \ and\ \bibinfo {author} {\bibfnamefont {P.~A.}\ \bibnamefont
  {McClarty}},\ }\href {\doibase 10.1103/PhysRevB.105.064430} {\bibfield
  {journal} {\bibinfo  {journal} {Phys. Rev. B}\ }\textbf {\bibinfo {volume}
  {105}},\ \bibinfo {pages} {064430} (\bibinfo {year} {2022})}\BibitemShut
  {NoStop}%
\bibitem [{Note1()}]{Note1}%
  \BibitemOpen
  \bibinfo {note} {This approximation requires that $\omega \ll D_1 l_0/d^2$.
  We nevertheless emphasize that the plots shown in the paper are for a finite
  $\omega = 1/\tau _s$.}\BibitemShut {Stop}%
\bibitem [{Note2()}]{Note2}%
  \BibitemOpen
  \bibinfo {note} {The singularity of the integrand at $\phi _k = 0 , \pi /2,
  ...$ and $D_2 = D_1$ is an artifact of the approximation $\omega \sim 0$ that
  was taken to write Eq.~\protect \eqref {eq:close}. The frequency $\omega $ is
  never zero, and thus the integrand is always finite. All plots shown in the
  paper are generated with the full expression including a non-vanishing
  frequency, which is shown in the SI.}\BibitemShut {Stop}%
\bibitem [{\citenamefont {Finco}\ and\ \citenamefont
  {Jacques}(2023)}]{Finco_2023_single_spin}%
  \BibitemOpen
  \bibfield  {author} {\bibinfo {author} {\bibfnamefont {A.}~\bibnamefont
  {Finco}}\ and\ \bibinfo {author} {\bibfnamefont {V.}~\bibnamefont
  {Jacques}},\ }\href {\doibase 10.1063/5.0167480} {\bibfield  {journal}
  {\bibinfo  {journal} {APL Materials}\ }\textbf {\bibinfo {volume} {11}},\
  \bibinfo {pages} {100901} (\bibinfo {year} {2023})}\BibitemShut {NoStop}%
\bibitem [{\citenamefont {Gugler}\ \emph {et~al.}(2018)\citenamefont {Gugler},
  \citenamefont {Astner}, \citenamefont {Angerer}, \citenamefont
  {Schmiedmayer}, \citenamefont {Majer},\ and\ \citenamefont
  {Mohn}}]{Gugler_2018Abinitio}%
  \BibitemOpen
  \bibfield  {author} {\bibinfo {author} {\bibfnamefont {J.}~\bibnamefont
  {Gugler}}, \bibinfo {author} {\bibfnamefont {T.}~\bibnamefont {Astner}},
  \bibinfo {author} {\bibfnamefont {A.}~\bibnamefont {Angerer}}, \bibinfo
  {author} {\bibfnamefont {J.}~\bibnamefont {Schmiedmayer}}, \bibinfo {author}
  {\bibfnamefont {J.}~\bibnamefont {Majer}}, \ and\ \bibinfo {author}
  {\bibfnamefont {P.}~\bibnamefont {Mohn}},\ }\href {\doibase
  10.1103/PhysRevB.98.214442} {\bibfield  {journal} {\bibinfo  {journal} {Phys.
  Rev. B}\ }\textbf {\bibinfo {volume} {98}},\ \bibinfo {pages} {214442}
  (\bibinfo {year} {2018})}\BibitemShut {NoStop}%
\bibitem [{\citenamefont {Norambuena}\ \emph {et~al.}(2018)\citenamefont
  {Norambuena}, \citenamefont {Mu\~noz}, \citenamefont {Dinani}, \citenamefont
  {Jarmola}, \citenamefont {Maletinsky}, \citenamefont {Budker},\ and\
  \citenamefont {Maze}}]{Norambuena2018spinlattice}%
  \BibitemOpen
  \bibfield  {author} {\bibinfo {author} {\bibfnamefont {A.}~\bibnamefont
  {Norambuena}}, \bibinfo {author} {\bibfnamefont {E.}~\bibnamefont {Mu\~noz}},
  \bibinfo {author} {\bibfnamefont {H.~T.}\ \bibnamefont {Dinani}}, \bibinfo
  {author} {\bibfnamefont {A.}~\bibnamefont {Jarmola}}, \bibinfo {author}
  {\bibfnamefont {P.}~\bibnamefont {Maletinsky}}, \bibinfo {author}
  {\bibfnamefont {D.}~\bibnamefont {Budker}}, \ and\ \bibinfo {author}
  {\bibfnamefont {J.~R.}\ \bibnamefont {Maze}},\ }\href {\doibase
  10.1103/PhysRevB.97.094304} {\bibfield  {journal} {\bibinfo  {journal} {Phys.
  Rev. B}\ }\textbf {\bibinfo {volume} {97}},\ \bibinfo {pages} {094304}
  (\bibinfo {year} {2018})}\BibitemShut {NoStop}%
\bibitem [{\citenamefont {Sangtawesin}\ \emph {et~al.}(2019)\citenamefont
  {Sangtawesin}, \citenamefont {Dwyer}, \citenamefont {Srinivasan},
  \citenamefont {Allred}, \citenamefont {Rodgers}, \citenamefont {De~Greve},
  \citenamefont {Stacey}, \citenamefont {Dontschuk}, \citenamefont {O'Donnell},
  \citenamefont {Hu}, \citenamefont {Evans}, \citenamefont {Jaye},
  \citenamefont {Fischer}, \citenamefont {Markham}, \citenamefont {Twitchen},
  \citenamefont {Park}, \citenamefont {Lukin},\ and\ \citenamefont
  {de~Leon}}]{Sangtawesin2019Origins}%
  \BibitemOpen
  \bibfield  {author} {\bibinfo {author} {\bibfnamefont {S.}~\bibnamefont
  {Sangtawesin}}, \bibinfo {author} {\bibfnamefont {B.~L.}\ \bibnamefont
  {Dwyer}}, \bibinfo {author} {\bibfnamefont {S.}~\bibnamefont {Srinivasan}},
  \bibinfo {author} {\bibfnamefont {J.~J.}\ \bibnamefont {Allred}}, \bibinfo
  {author} {\bibfnamefont {L.~V.~H.}\ \bibnamefont {Rodgers}}, \bibinfo
  {author} {\bibfnamefont {K.}~\bibnamefont {De~Greve}}, \bibinfo {author}
  {\bibfnamefont {A.}~\bibnamefont {Stacey}}, \bibinfo {author} {\bibfnamefont
  {N.}~\bibnamefont {Dontschuk}}, \bibinfo {author} {\bibfnamefont {K.~M.}\
  \bibnamefont {O'Donnell}}, \bibinfo {author} {\bibfnamefont {D.}~\bibnamefont
  {Hu}}, \bibinfo {author} {\bibfnamefont {D.~A.}\ \bibnamefont {Evans}},
  \bibinfo {author} {\bibfnamefont {C.}~\bibnamefont {Jaye}}, \bibinfo {author}
  {\bibfnamefont {D.~A.}\ \bibnamefont {Fischer}}, \bibinfo {author}
  {\bibfnamefont {M.~L.}\ \bibnamefont {Markham}}, \bibinfo {author}
  {\bibfnamefont {D.~J.}\ \bibnamefont {Twitchen}}, \bibinfo {author}
  {\bibfnamefont {H.}~\bibnamefont {Park}}, \bibinfo {author} {\bibfnamefont
  {M.~D.}\ \bibnamefont {Lukin}}, \ and\ \bibinfo {author} {\bibfnamefont
  {N.~P.}\ \bibnamefont {de~Leon}},\ }\href {\doibase
  10.1103/PhysRevX.9.031052} {\bibfield  {journal} {\bibinfo  {journal} {Phys.
  Rev. X}\ }\textbf {\bibinfo {volume} {9}},\ \bibinfo {pages} {031052}
  (\bibinfo {year} {2019})}\BibitemShut {NoStop}%
\bibitem [{\citenamefont {Candido}\ and\ \citenamefont
  {Flatt\'e}(2024)}]{candido2024interplay}%
  \BibitemOpen
  \bibfield  {author} {\bibinfo {author} {\bibfnamefont {D.~R.}\ \bibnamefont
  {Candido}}\ and\ \bibinfo {author} {\bibfnamefont {M.~E.}\ \bibnamefont
  {Flatt\'e}},\ }\href {\doibase 10.1103/PhysRevB.110.024419} {\bibfield
  {journal} {\bibinfo  {journal} {Phys. Rev. B}\ }\textbf {\bibinfo {volume}
  {110}},\ \bibinfo {pages} {024419} (\bibinfo {year} {2024})}\BibitemShut
  {NoStop}%
\bibitem [{\citenamefont {Irber}\ \emph {et~al.}(2021)\citenamefont {Irber},
  \citenamefont {Poggiali}, \citenamefont {Kong}, \citenamefont {Kieschnick},
  \citenamefont {L\"{u}hmann}, \citenamefont {Kwiatkowski}, \citenamefont
  {Meijer}, \citenamefont {Du}, \citenamefont {Shi},\ and\ \citenamefont
  {Reinhard}}]{Irber2021RobustallOptical}%
  \BibitemOpen
  \bibfield  {author} {\bibinfo {author} {\bibfnamefont {D.~M.}\ \bibnamefont
  {Irber}}, \bibinfo {author} {\bibfnamefont {F.}~\bibnamefont {Poggiali}},
  \bibinfo {author} {\bibfnamefont {F.}~\bibnamefont {Kong}}, \bibinfo {author}
  {\bibfnamefont {M.}~\bibnamefont {Kieschnick}}, \bibinfo {author}
  {\bibfnamefont {T.}~\bibnamefont {L\"{u}hmann}}, \bibinfo {author}
  {\bibfnamefont {D.}~\bibnamefont {Kwiatkowski}}, \bibinfo {author}
  {\bibfnamefont {J.}~\bibnamefont {Meijer}}, \bibinfo {author} {\bibfnamefont
  {J.}~\bibnamefont {Du}}, \bibinfo {author} {\bibfnamefont {F.}~\bibnamefont
  {Shi}}, \ and\ \bibinfo {author} {\bibfnamefont {F.}~\bibnamefont
  {Reinhard}},\ }\href {\doibase 10.1038/s41467-020-20755-3} {\bibfield
  {journal} {\bibinfo  {journal} {Nature Communications}\ }\textbf {\bibinfo
  {volume} {12}} (\bibinfo {year} {2021}),\
  10.1038/s41467-020-20755-3}\BibitemShut {NoStop}%
\bibitem [{\citenamefont {Cambria}\ \emph {et~al.}(2023)\citenamefont
  {Cambria}, \citenamefont {Norambuena}, \citenamefont {Dinani}, \citenamefont
  {Thiering}, \citenamefont {Gardill}, \citenamefont {Kemeny}, \citenamefont
  {Li}, \citenamefont {Lordi}, \citenamefont {Gali}, \citenamefont {Maze},\
  and\ \citenamefont {Kolkowitz}}]{cambria_2023_temperature_dependent}%
  \BibitemOpen
  \bibfield  {author} {\bibinfo {author} {\bibfnamefont {M.~C.}\ \bibnamefont
  {Cambria}}, \bibinfo {author} {\bibfnamefont {A.}~\bibnamefont {Norambuena}},
  \bibinfo {author} {\bibfnamefont {H.~T.}\ \bibnamefont {Dinani}}, \bibinfo
  {author} {\bibfnamefont {G.}~\bibnamefont {Thiering}}, \bibinfo {author}
  {\bibfnamefont {A.}~\bibnamefont {Gardill}}, \bibinfo {author} {\bibfnamefont
  {I.}~\bibnamefont {Kemeny}}, \bibinfo {author} {\bibfnamefont
  {Y.}~\bibnamefont {Li}}, \bibinfo {author} {\bibfnamefont {V.}~\bibnamefont
  {Lordi}}, \bibinfo {author} {\bibfnamefont {A.}~\bibnamefont {Gali}},
  \bibinfo {author} {\bibfnamefont {J.~R.}\ \bibnamefont {Maze}}, \ and\
  \bibinfo {author} {\bibfnamefont {S.}~\bibnamefont {Kolkowitz}},\ }\href
  {\doibase 10.1103/PhysRevLett.130.256903} {\bibfield  {journal} {\bibinfo
  {journal} {Phys. Rev. Lett.}\ }\textbf {\bibinfo {volume} {130}},\ \bibinfo
  {pages} {256903} (\bibinfo {year} {2023})}\BibitemShut {NoStop}%
\bibitem [{\citenamefont {Weggler}\ \emph {et~al.}(2020)\citenamefont
  {Weggler}, \citenamefont {Ganslmayer}, \citenamefont {Frank}, \citenamefont
  {Eilert}, \citenamefont {Jelezko},\ and\ \citenamefont
  {Michaelis}}]{Weggler2020Determination}%
  \BibitemOpen
  \bibfield  {author} {\bibinfo {author} {\bibfnamefont {T.}~\bibnamefont
  {Weggler}}, \bibinfo {author} {\bibfnamefont {C.}~\bibnamefont {Ganslmayer}},
  \bibinfo {author} {\bibfnamefont {F.}~\bibnamefont {Frank}}, \bibinfo
  {author} {\bibfnamefont {T.}~\bibnamefont {Eilert}}, \bibinfo {author}
  {\bibfnamefont {F.}~\bibnamefont {Jelezko}}, \ and\ \bibinfo {author}
  {\bibfnamefont {J.}~\bibnamefont {Michaelis}},\ }\href {\doibase
  10.1021/acs.nanolett.9b04725} {\bibfield  {journal} {\bibinfo  {journal}
  {Nano Letters}\ }\textbf {\bibinfo {volume} {20}},\ \bibinfo {pages} {2980}
  (\bibinfo {year} {2020})}\BibitemShut {NoStop}%
\bibitem [{\citenamefont {Edmonds}\ \emph {et~al.}(2012)\citenamefont
  {Edmonds}, \citenamefont {D'Haenens-Johansson}, \citenamefont {Cruddace},
  \citenamefont {Newton}, \citenamefont {Fu}, \citenamefont {Santori},
  \citenamefont {Beausoleil}, \citenamefont {Twitchen},\ and\ \citenamefont
  {Markham}}]{Edmonds2012Production}%
  \BibitemOpen
  \bibfield  {author} {\bibinfo {author} {\bibfnamefont {A.~M.}\ \bibnamefont
  {Edmonds}}, \bibinfo {author} {\bibfnamefont {U.~F.~S.}\ \bibnamefont
  {D'Haenens-Johansson}}, \bibinfo {author} {\bibfnamefont {R.~J.}\
  \bibnamefont {Cruddace}}, \bibinfo {author} {\bibfnamefont {M.~E.}\
  \bibnamefont {Newton}}, \bibinfo {author} {\bibfnamefont {K.-M.~C.}\
  \bibnamefont {Fu}}, \bibinfo {author} {\bibfnamefont {C.}~\bibnamefont
  {Santori}}, \bibinfo {author} {\bibfnamefont {R.~G.}\ \bibnamefont
  {Beausoleil}}, \bibinfo {author} {\bibfnamefont {D.~J.}\ \bibnamefont
  {Twitchen}}, \ and\ \bibinfo {author} {\bibfnamefont {M.~L.}\ \bibnamefont
  {Markham}},\ }\href {\doibase 10.1103/PhysRevB.86.035201} {\bibfield
  {journal} {\bibinfo  {journal} {Phys. Rev. B}\ }\textbf {\bibinfo {volume}
  {86}},\ \bibinfo {pages} {035201} (\bibinfo {year} {2012})}\BibitemShut
  {NoStop}%
\bibitem [{\citenamefont {Klink}\ \emph {et~al.}(2025)\citenamefont {Klink},
  \citenamefont {Kirkpatrick}, \citenamefont {Tadokoro}, \citenamefont
  {Becker},\ and\ \citenamefont {Nicley}}]{Klink2025Fabrication}%
  \BibitemOpen
  \bibfield  {author} {\bibinfo {author} {\bibfnamefont {K.}~\bibnamefont
  {Klink}}, \bibinfo {author} {\bibfnamefont {A.~R.}\ \bibnamefont
  {Kirkpatrick}}, \bibinfo {author} {\bibfnamefont {Y.}~\bibnamefont
  {Tadokoro}}, \bibinfo {author} {\bibfnamefont {J.~N.}\ \bibnamefont
  {Becker}}, \ and\ \bibinfo {author} {\bibfnamefont {S.~S.}\ \bibnamefont
  {Nicley}},\ }\href {\doibase 10.3389/frqst.2025.1667545} {\bibfield
  {journal} {\bibinfo  {journal} {Frontiers in Quantum Science and Technology}\
  }\textbf {\bibinfo {volume} {4}} (\bibinfo {year} {2025}),\
  10.3389/frqst.2025.1667545}\BibitemShut {NoStop}%
\bibitem [{\citenamefont {Cardoso~Barbosa}\ \emph {et~al.}(2024)\citenamefont
  {Cardoso~Barbosa}, \citenamefont {Gutsche}, \citenamefont {L\"onard},
  \citenamefont {Dix},\ and\ \citenamefont {Widera}}]{Barbosa2024Temperature}%
  \BibitemOpen
  \bibfield  {author} {\bibinfo {author} {\bibfnamefont {I.}~\bibnamefont
  {Cardoso~Barbosa}}, \bibinfo {author} {\bibfnamefont {J.}~\bibnamefont
  {Gutsche}}, \bibinfo {author} {\bibfnamefont {D.}~\bibnamefont {L\"onard}},
  \bibinfo {author} {\bibfnamefont {S.}~\bibnamefont {Dix}}, \ and\ \bibinfo
  {author} {\bibfnamefont {A.}~\bibnamefont {Widera}},\ }\href {\doibase
  10.1103/PhysRevResearch.6.023078} {\bibfield  {journal} {\bibinfo  {journal}
  {Phys. Rev. Res.}\ }\textbf {\bibinfo {volume} {6}},\ \bibinfo {pages}
  {023078} (\bibinfo {year} {2024})}\BibitemShut {NoStop}%
\bibitem [{\citenamefont {Chatterjee}\ \emph {et~al.}(2022)\citenamefont
  {Chatterjee}, \citenamefont {Dolgirev}, \citenamefont {Esterlis},
  \citenamefont {Zibrov}, \citenamefont {Lukin}, \citenamefont {Yao},\ and\
  \citenamefont {Demler}}]{Chatterjee_2Dsupercond2022}%
  \BibitemOpen
  \bibfield  {author} {\bibinfo {author} {\bibfnamefont {S.}~\bibnamefont
  {Chatterjee}}, \bibinfo {author} {\bibfnamefont {P.~E.}\ \bibnamefont
  {Dolgirev}}, \bibinfo {author} {\bibfnamefont {I.}~\bibnamefont {Esterlis}},
  \bibinfo {author} {\bibfnamefont {A.~A.}\ \bibnamefont {Zibrov}}, \bibinfo
  {author} {\bibfnamefont {M.~D.}\ \bibnamefont {Lukin}}, \bibinfo {author}
  {\bibfnamefont {N.~Y.}\ \bibnamefont {Yao}}, \ and\ \bibinfo {author}
  {\bibfnamefont {E.}~\bibnamefont {Demler}},\ }\href {\doibase
  10.1103/PhysRevResearch.4.L012001} {\bibfield  {journal} {\bibinfo  {journal}
  {Phys. Rev. Res.}\ }\textbf {\bibinfo {volume} {4}},\ \bibinfo {pages}
  {L012001} (\bibinfo {year} {2022})}\BibitemShut {NoStop}%
\bibitem [{\citenamefont {Machado}\ \emph {et~al.}(2023)\citenamefont
  {Machado}, \citenamefont {Demler}, \citenamefont {Yao},\ and\ \citenamefont
  {Chatterjee}}]{machado_2023_quantum_noise}%
  \BibitemOpen
  \bibfield  {author} {\bibinfo {author} {\bibfnamefont {F.}~\bibnamefont
  {Machado}}, \bibinfo {author} {\bibfnamefont {E.~A.}\ \bibnamefont {Demler}},
  \bibinfo {author} {\bibfnamefont {N.~Y.}\ \bibnamefont {Yao}}, \ and\
  \bibinfo {author} {\bibfnamefont {S.}~\bibnamefont {Chatterjee}},\ }\href
  {\doibase 10.1103/PhysRevLett.131.070801} {\bibfield  {journal} {\bibinfo
  {journal} {Phys. Rev. Lett.}\ }\textbf {\bibinfo {volume} {131}},\ \bibinfo
  {pages} {070801} (\bibinfo {year} {2023})}\BibitemShut {NoStop}%
\bibitem [{\citenamefont {Rovny}\ \emph {et~al.}(2022)\citenamefont {Rovny},
  \citenamefont {Yuan}, \citenamefont {Fitzpatrick}, \citenamefont {Abdalla},
  \citenamefont {Futamura}, \citenamefont {Fox}, \citenamefont {Cambria},
  \citenamefont {Kolkowitz},\ and\ \citenamefont
  {de~Leon}}]{rovny2022nanoscale}%
  \BibitemOpen
  \bibfield  {author} {\bibinfo {author} {\bibfnamefont {J.}~\bibnamefont
  {Rovny}}, \bibinfo {author} {\bibfnamefont {Z.}~\bibnamefont {Yuan}},
  \bibinfo {author} {\bibfnamefont {M.}~\bibnamefont {Fitzpatrick}}, \bibinfo
  {author} {\bibfnamefont {A.~I.}\ \bibnamefont {Abdalla}}, \bibinfo {author}
  {\bibfnamefont {L.}~\bibnamefont {Futamura}}, \bibinfo {author}
  {\bibfnamefont {C.}~\bibnamefont {Fox}}, \bibinfo {author} {\bibfnamefont
  {M.~C.}\ \bibnamefont {Cambria}}, \bibinfo {author} {\bibfnamefont
  {S.}~\bibnamefont {Kolkowitz}}, \ and\ \bibinfo {author} {\bibfnamefont
  {N.~P.}\ \bibnamefont {de~Leon}},\ }\href {\doibase 10.1126/science.ade9858}
  {\bibfield  {journal} {\bibinfo  {journal} {Science}\ }\textbf {\bibinfo
  {volume} {378}},\ \bibinfo {pages} {1301} (\bibinfo {year}
  {2022})}\BibitemShut {NoStop}%
\bibitem [{\citenamefont {Hosseinabadi}\ \emph {et~al.}(2025)\citenamefont
  {Hosseinabadi} \emph {et~al.}}]{Hosseinabadi_twoNV2025}%
  \BibitemOpen
  \bibfield  {author} {\bibinfo {author} {\bibfnamefont {H.}~\bibnamefont
  {Hosseinabadi}} \emph {et~al.},\ }\href@noop {} {\bibfield  {journal}
  {\bibinfo  {journal} {Under preparation}\ } (\bibinfo {year}
  {2025})}\BibitemShut {NoStop}%
\bibitem [{\citenamefont {Fang}\ \emph {et~al.}(2022)\citenamefont {Fang},
  \citenamefont {Zhang},\ and\ \citenamefont
  {Tserkovnyak}}]{fang_2022_generalizedmodel}%
  \BibitemOpen
  \bibfield  {author} {\bibinfo {author} {\bibfnamefont {H.}~\bibnamefont
  {Fang}}, \bibinfo {author} {\bibfnamefont {S.}~\bibnamefont {Zhang}}, \ and\
  \bibinfo {author} {\bibfnamefont {Y.}~\bibnamefont {Tserkovnyak}},\ }\href
  {\doibase 10.1103/PhysRevB.105.184406} {\bibfield  {journal} {\bibinfo
  {journal} {Phys. Rev. B}\ }\textbf {\bibinfo {volume} {105}},\ \bibinfo
  {pages} {184406} (\bibinfo {year} {2022})}\BibitemShut {NoStop}%
\bibitem [{\citenamefont {Flebus}(2019)}]{flebus_2019_chemical_potential}%
  \BibitemOpen
  \bibfield  {author} {\bibinfo {author} {\bibfnamefont {B.}~\bibnamefont
  {Flebus}},\ }\href {\doibase 10.1103/PhysRevB.100.064410} {\bibfield
  {journal} {\bibinfo  {journal} {Phys. Rev. B}\ }\textbf {\bibinfo {volume}
  {100}},\ \bibinfo {pages} {064410} (\bibinfo {year} {2019})}\BibitemShut
  {NoStop}%
\end{thebibliography}%

\end{document}